\documentclass[pdflatex,sn-mathphys-num]{sn-jnl}
\usepackage{makecell}
\usepackage{graphicx}%
\usepackage{multirow}%
\usepackage{amsmath,amssymb,amsfonts}%
\usepackage{amsthm}%
\usepackage{mathrsfs}%
\usepackage[title]{appendix}%
\usepackage{xcolor}%
\usepackage{textcomp}%
\usepackage{manyfoot}%
\usepackage{booktabs}%
\usepackage{algorithm}%
\usepackage{algorithmicx}%
\usepackage{algpseudocode}%
\usepackage{listings}%
\usepackage{diagbox}

\usepackage{xcolor}

\definecolor{rblue}{rgb}{0,0.5,1}
\definecolor{awesome}{rgb}{1.0, 0.13, 0.32}
\definecolor{gy}{rgb}{0.5, 0, 0.5}
\def\eg{\textit{e.g.}} 
\def\ie{\textit{i.e}.}

 \def\vs{\textit{vs}.}

\definecolor{hollywoodcerise}{rgb}{0.96, 0.0, 0.63}
\definecolor{lasallegreen}{rgb}{0.03, 0.47, 0.19}
\definecolor{hanpurple}{rgb}{0.32, 0.09, 0.98}
\definecolor{green(pigment)}{rgb}{0.0, 0.65, 0.31}

\theoremstyle{thmstyleone}%
\theoremstyle{thmstyletwo}%

\theoremstyle{thmstylethree}%

\begin{document}

\title[Article Title]{Neuro-inspired automated lens design}

\author[1]{\sur{Yao Gao}}\email{gaoyao@zju.edu.cn}

\author*[1,2,]{\sur{Lei Sun}}\email{leo\_sun@zju.edu.cn}

\author[1]{\sur{Shaohua Gao}}\email{gaoshaohua@zju.edu.cn}

\author[1]{\sur{Qi Jiang}}\email{qijiang@zju.edu.cn}

\author[3]{\sur{Kailun Yang}}\email{kailun.yang@hnu.edu.cn}

\author[1]{\sur{Weijian Hu}}\email{huweijian@zju.edu.cn}

\author[1]{\sur{Xiaolong Qian}}\email{xiaolongqian@zju.edu.cn}

\author[1]{\sur{Wenyong Li}}\email{liwenyong@zju.edu.cn}

\author[2,4]{\sur{Luc Van Gool}}\email{vangool@vision.ee.ethz.ch}

\author*[1,]{\sur{Kaiwei Wang}}\email{wangkaiwei@zju.edu.cn}

\affil*[1]{\orgdiv{Zhejiang University}, \orgname{State Key Laboratory of Extreme Photonics and Instrumentation}, \orgaddress{\city{Hangzhou}, \postcode{310027}, \country{China}}}

\affil[2]{\orgdiv{INSAIT}, \orgname{Sofia University St. ``Kliment Ohridski''}, \orgaddress{\city{Sofia}, \postcode{1784}, \country{Bulgaria}}}

\affil[3]{\orgdiv{Hunan University}, \orgname{School of Artificial Intelligence and Robotics}, \orgaddress{\city{Changsha}, \postcode{410012}, \country{China}}}

\affil[4]{\orgdiv{ETH Zurich}, \orgname{Computer Vision Lab}, \orgaddress{\city{Zurich}, \postcode{8092}, \country{Switzerland}}}

\abstract{
The highly non-convex optimization landscape of modern lens design necessitates extensive human expertise, resulting in inefficiency and constrained design diversity. While automated methods are desirable, existing approaches remain limited to simple tasks or produce complex lenses with suboptimal image quality.
Drawing inspiration from the synaptic pruning mechanism in mammalian neural development, this study proposes OptiNeuro—a novel automated lens design framework that first generates diverse initial structures and 
then progressively eliminates low-performance lenses while refining remaining candidates through gradient-based optimization. 
By fully automating the design of complex aspheric imaging lenses, OptiNeuro demonstrates quasi-human-level performance, identifying multiple viable candidates with minimal human intervention. This advancement not only enhances the automation level and efficiency of lens design but also facilitates the exploration of previously uncharted lens architectures.
}

\maketitle

\section{Introduction}

Lens design, the systematic process of initializing and optimizing lens parameters to meet desired image quality under certain physical constraints, has been a topic of considerable importance in advancing how humans explore the unknown world~\cite{wang2015witnessing, gissibl2016sub, zhang2023large}. 
Since lens design is a daunting task, automated lens design involving no or minimal human effort has always been the expectation of scientists, researchers, and optical engineers~\cite{yang2017automated}.
Today, the development of several commercial lens design software such as Zemax and CODE V has dramatically simplified the lens design process~\cite{laikin2018lens}. However, these software still require extensive manual intervention to achieve complex lens design. 
Subsequent research has sought to further mitigate human dependency in lens design through diverse computational approaches, including heuristic search algorithms~\cite{guo2019new, zhang2020automated, gao2025exploring}, Deep Neural Networks (DNNs)~\cite{cote2021deep}, point-by-point optimization~\cite{yang2017automated, zhang2021towards}, and curriculum learning frameworks~\cite{yang2024curriculum}. However, these works remain limited in scope and performance. They are typically restricted to simple configurations (\eg, Cooke triplets, Double Gauss lenses), specific design types (\eg, freeform lenses), or produce complex systems (\eg, smartphone lenses) with suboptimal image quality (\textcolor{blue}{Supplementary Note~1}). 

Therefore, the automated design of complex lenses remains a formidable challenge.
The primary reason lies in the highly non-convex optimization landscape in lens design, which makes it prone to converging to suboptimal local minima that are widely distributed within the solution space~\cite{yang2024curriculum, sun2021end}. 
Without manual establishment of suitable initial structures and adjustment of lens parameters to escape suboptimal local minima, such designs risk overlooking superior alternatives~\cite{sun2016lens}. 
A preliminary idea involves generating large-scale optimization starting points; however, given constrained computational resources, optimizing an excessive number of candidates leads to prohibitively low efficiency. Consequently, the problem transforms into identifying high-quality solutions from a vast pool of candidates while operating within computational limits.

\begin{figure}[!t]
    \centering
    \includegraphics[width=1\linewidth]{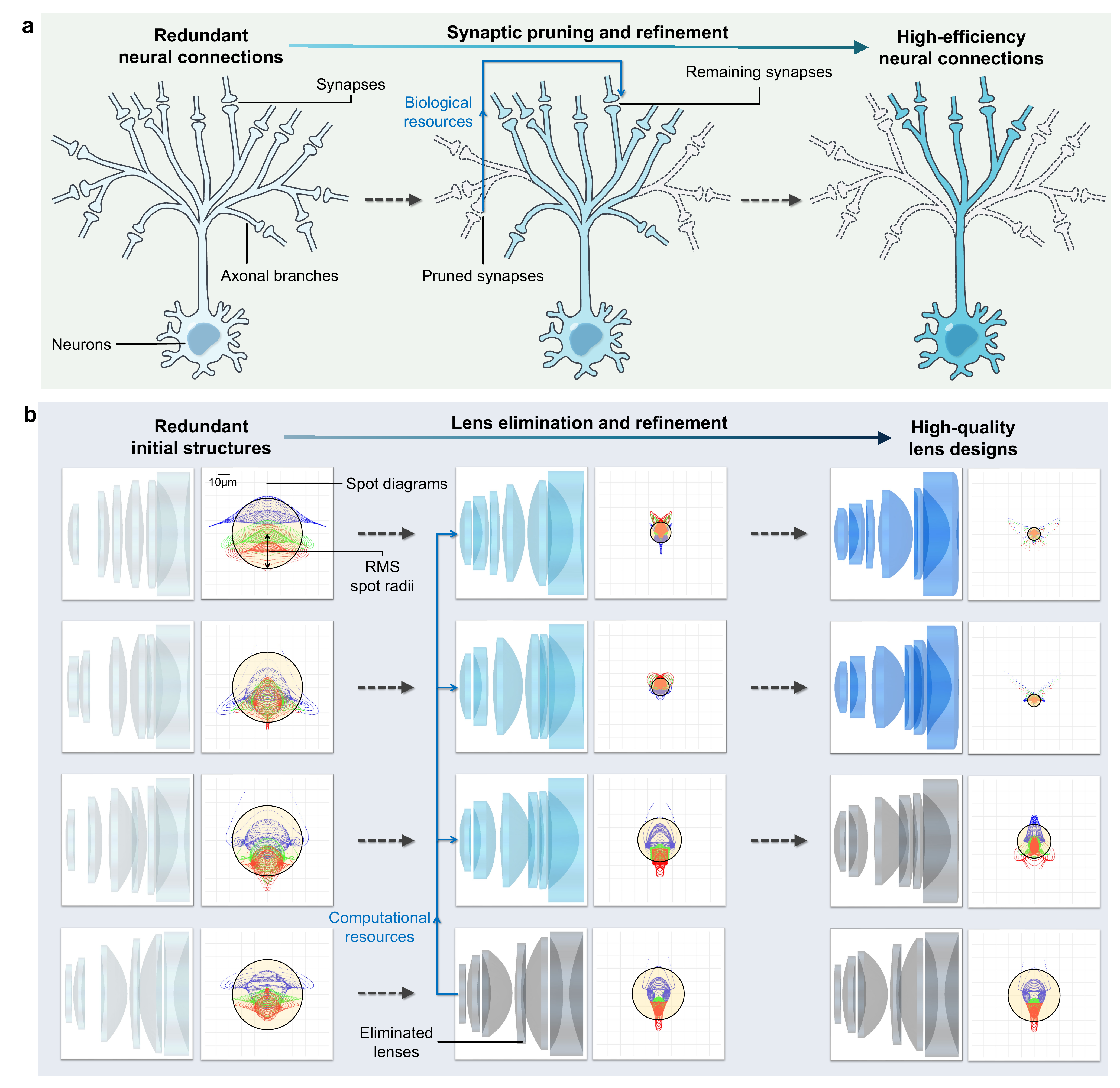}
    \caption{\textbf{Automated lens design inspired by the synaptic pruning mechanism in mammalian neural development.} 
    \textbf{a} Developing mammalian brains establish redundant neural connections through synaptic overgrowth during early development. These connections are initially imprecise and become more precise through synaptic pruning, which involves weakening and eliminating some synaptic connections and strengthening others. 
    Ultimately, high-efficiency and low-redundancy neural connections are formed. 
    \textbf{b} Drawing from this biological principle, OptiNeuro generates a group of redundant initial structures given a set of design specifications. During the iterative optimization, low-quality lenses are progressively eliminated.
    Ultimately, a set of high-quality lens designs is obtained.
    Here, we employ spot diagrams as a critical metric for evaluating lens quality. A smaller RMS (Root Mean Square) spot radius typically indicates superior imaging performance.}
    \label{fig:intro}
\end{figure}

Mammalian brain development faces a parallel challenge: how to sculpt efficient neural circuits using limited biological resources. To address this, developing mammalian brains first establish redundant and imprecise neural connections through a phase of synaptic proliferation during early development~\cite{somaiya2024visualizing}, followed by a systematic process of synaptic pruning that progressively eliminates underutilized synapses, as illustrated in Fig.~\ref{fig:intro}a. Simultaneously, limited biological resources are strategically reallocated to reinforce and maintain more relevant and frequently activated connections~\cite{faust2021mechanisms}, ultimately yielding optimized and functionally specialized neural circuitry.
Leveraging this biological paradigm, OptiNeuro adopts an initial set of redundant candidate structures. As shown in Fig.~\ref{fig:intro}b, the framework iteratively eliminates low-performance lenses while dynamically reallocating computational resources to refine remaining candidates, ultimately producing a set of high-quality lenses that meet the specified requirements. 
It also comes with an incremental optimization strategy to reduce the optimization complexity in early design stages and a random perturbation strategy to escape local minima. 

We validated OptiNeuro's effectiveness through three progressive validation frameworks: 
(1) a six-element aspheric lens design task illustrating improved performance over the state-of-the-art automated lens design method based on curriculum learning (Sec.~\ref{sec:2_1}), 
(2) four nine-element aspheric lens design tasks demonstrating its quasi-human-level design capabilities (Sec.~\ref{sec:2_2}), and 
(3) a novel glass-plastic hybrid fisheye lens design task highlighting its potential for facilitating the exploration of uncharted lens architectures (Sec.~\ref{sec:2_3}). We believe that OptiNeuro allows designers to be relieved from burdensome tasks of searching for suitable initial structures and manually optimizing lens quality, and shift their focus toward evaluating high-quality candidate solutions and applying targeted refinements. 
More significantly, OptiNeuro holds the potential to continuously expand the scale of public lens databases and enhance lens design efficiency, thereby exerting a transformative impact on the field of lens design.

\section{Results}
In this work, lens design is mathematically formulated as an optimization problem where the Merit Function—composed of an image quality function and a physical constraint function—is minimized (``Methods''). 
Given a set of design specifications, OptiNeuro first generates $N^{(0)}_L$ initial lens structures by employing a physics-constrained automated initialization strategy (``Methods''). 
The subsequent design process involves $K$ steps of lens elimination and refinement.
In the $i_{th}$ step ($i=1,2,...,K$), the number of remaining lenses is denoted as $N^{(i)}_L$, and $N^{(i-1)}_L-N^{(i)}_L$ lenses with inferior Merit Function performance are eliminated. 
To fully utilize the computational resources released by the eliminated lenses, OptiNeuro incorporates an incremental optimization strategy that dynamically reallocates these resources to the remaining candidates, expanding their variable spaces. 
Specifically, we set the number of variables for each intermediate step as:
\begin{equation}
\label{eq:result1}
N^{(i)}_V=\Big \lfloor \frac{C}{N^{(i)}_L} \Big \rfloor=N^{(0)}_V+(N^{(K)}_V-N^{(0)}_V)\times \frac{i}{K} \quad (i = 1, 2, \ldots, K),
\end{equation}
where $C$ is defined as a constant representing the total computational resources, $\lfloor \cdot \rfloor$ represents the floor function, and $N^{(0)}_V$ and $N^{(K)}_V$ are the initial and final number of variables, respectively. 
After lens elimination and computational resource reallocation, the following refinement process comprises $k$ cycles of random perturbations (``Methods'') and Adam-based local optimizations (``Methods''). Finally, $N^{(K)}_L$ high-quality lenses are obtained.

The most widely used lenses are spherical/aspheric lenses. We demonstrate that OptiNeuro can achieve automated design of multi-element spherical lenses in \textcolor{blue}{Supplementary Note~4}.
Compared to simple spherical lenses, aspheric lenses typically involve higher optimization complexity and greater design challenges. 
Therefore, we validate OptiNeuro's effectiveness through automated design of various aspheric lenses in the subsequent sections.

\begin{figure}
    \centering
    \includegraphics[width=1\linewidth]{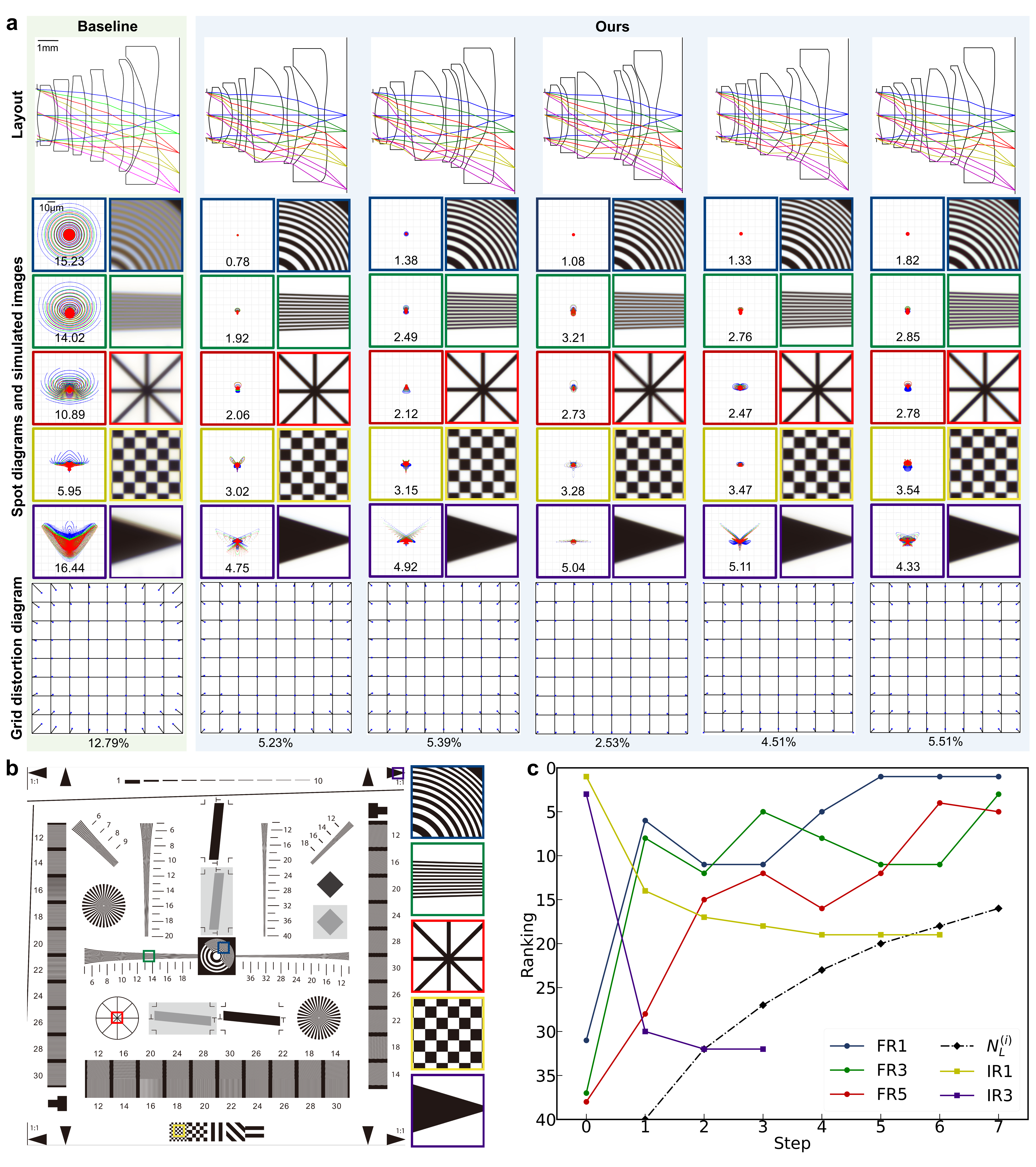}
    \caption{\textbf{Automated design of a six-element aspheric lens.} 
    \textbf{a} The baseline design and top-$5$ lenses designed by OptiNeuro.
    For each involved lens, its layout, spot diagrams across five sampled fields (with corresponding RMS spot radii labeled below and zoomed patches of simulated images illustrated on the right), and grid distortion diagram (annotated with maximum distortion value below) are illustrated from top to bottom. 
    \textbf{b} ISO 12233 resolution chart and five zoomed patches corresponding to the five sampled field positions.
    \textbf{c} The ranking progression of five selected lenses (labeled as FR1, FR3, FR5, IR1 and IR3) throughout entire $7$ steps of lens elimination and refinement.
    Lenses with a final ranking of $j$ are labeled as FR$j$, and those with an initial ranking of $j$ are labeled as IR$j$. $N^{(i)}_L$ represents the number of remaining lenses in the $i_{th}$ step.
    }  
    \label{fig:phone6}
\end{figure}

\subsection{Automated design of a six-element aspheric lens}
\label{sec:2_1}
The state-of-the-art automated lens design method based on curriculum learning has been successfully applied to a six-element aspheric lens~\cite{yang2024curriculum}, with specifications including a Field OF View (FOV) of $80.8^\circ$, an F-number of $2.0$, and a sensor diagonal length of $7.66 mm$; therefore, we showcased an example of automatically designing this lens. In this design task, the average (Avg) RMS spot radius across five uniformly sampled fields and three standard wavelengths (F-, d-, and C-line wavelengths) was adopted as the image quality metric.
Building upon the original design specifications, we incorporated stringent constraints on critical physical quantities such as Effective Focal Length (EFL, $4.5mm$), Total Track Length (TTL, $<6.94mm$), Back Focal Length (BFL, $>1mm$), distortion($<6\%$), element thickness ($>0.3mm$), air spacing ($>0.1mm$), ray incident angle ($<60^\circ$), and surface slope angle ($<45^\circ$), thereby enhancing the rationality of design outcomes (\textcolor{blue}{Supplementary Note~3.1}). 

This design task was conducted on an NVIDIA GeForce RTX 4090 GPU, which limited $C$ in Eq.~\ref{eq:result1} to $2000$. 
Initially, only basic parameters were set as variables, including curvatures ($c$), distances ($d$), and conic coefficients ($\kappa$). 
In the $i_{th}$ step, higher-order aspheric coefficients ($a_{2(i+1)}$) were newly defined as variables, and the maximum aspheric coefficient was set to $a_{16}$ according to the definition of standard aspheric surfaces (``Methods''). Correspondingly, $N^{(0)}_V$,  $N^{(K)}_V$, and $K$ in Eq.~\ref{eq:result1} were set to $37$, $121$, and $7$ respectively. In addition, $k$ was set to $20$ in the refinement process to ensure the lens performance was fully optimized.
Under such settings, this design task was completed within three days, $54$ lens structures were initially generated and $16$ high-quality lens designs were finally obtained.

We selected the top-$5$ lenses from the $16$ candidates based on their Merit Function Values (MFVs) and used the design result in work~\cite{yang2024curriculum} as the baseline. 
To ensure fairness in comparative analysis, all design outcomes were imported into Zemax to analyze their spot diagrams and grid distortion diagrams. 
To provide a visual comparison of image quality, the simulation results of these lenses on the ISO 12233 resolution chart (a standard test pattern for evaluating imaging systems) were further calculated and five zoomed patches corresponding to the sampled field positions are selected for illustration (see Fig.~\ref{fig:phone6}b for the ground truth and refer to work~\cite{yang2024curriculum} for details of the simulation process).
For each involved lens, Fig.~\ref{fig:phone6}a illustrates its layout, spot diagrams across five sampled fields (with corresponding RMS spot radii labeled below and zoomed patches of simulated images illustrated on the right), and grid distortion diagram (annotated with maximum distortion value below). These experimental results demonstrate that OptiNeuro-designed lenses reduce Avg RMS spot radius from $12.506\ {\mu}m$ (baseline) to $2.506\sim3.064\ {\mu}m$ while maintaining distortion $<6\%$ (\vs $12.79\%$ of the baseline), confirming enhanced imaging performance.

We selected three final solutions and two mid-process eliminated lenses to demonstrate OptiNeuro's effectiveness. Fig.~\ref{fig:phone6}c illustrates the ranking progression through the seven-step elimination process. Lenses achieving final ranking $j$ are labeled FR$j$, while those initially ranked $j$ are denoted IR$j$. 
The ranking evolution exhibited significant unpredictability: (1) FR1, FR3, and FR5 initially ranked outside the top $30$ but ultimately achieved top-$5$ positions, and (2) IR1 and IR3 started within the top $5$ but were eliminated in steps $6$ and $3$, respectively. This contrast highlights a critical insight: Selecting only a single promising candidate (\eg, IR1) from a non-diverse initial set risks overlooking superior solutions. OptiNeuro mitigates this limitation through initial pool diversification, thereby improving solution quality.

\subsection{Automated design of four nine-element aspheric lenses}
\label{sec:2_2}
We further tested OptiNeuro's design capabilities on more complex aspheric lens tasks. 
Based on existing public patent data~\cite{chen2022optical, chen2024imaging}, we established design specifications for four nine-element aspheric lenses and constructed reference manual designs (\textcolor{blue}{Supplementary Note~5.1}). 
Following manual design results, the highest aspheric coefficient was set to $a_{20}$, resulting in $N^{(0)}_V=54$, $N^{(K)}_V=216$, and $K=9$ in Eq.~\ref{eq:result1}. All other experimental parameters remained consistent with previous experiments.

\begin{figure}
    \centering
    \includegraphics[width=0.98\linewidth]{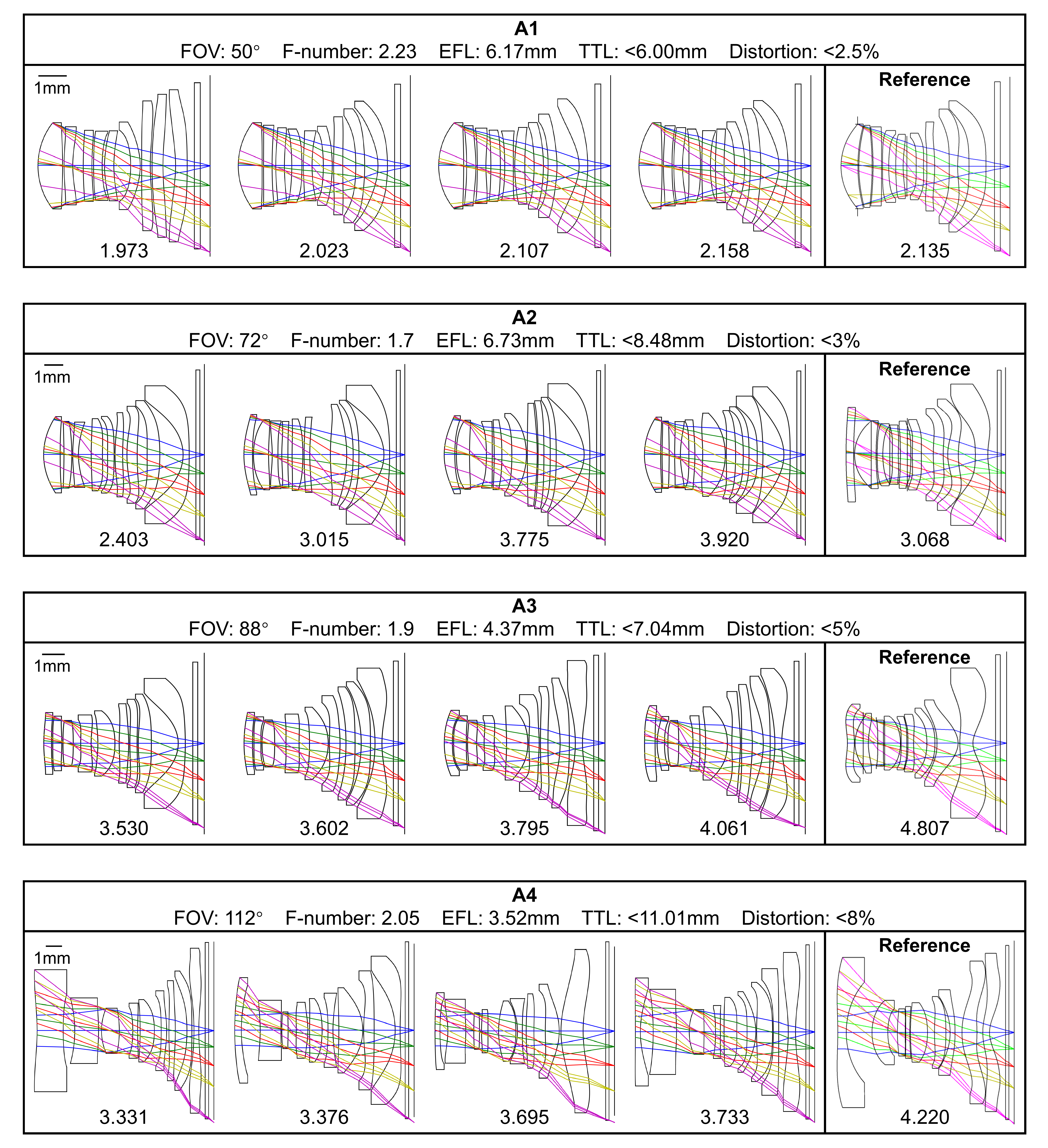}
    \caption{\textbf{Design results of nine-element aspheric lenses.} Four sets of design specifications are labeled respectively as A1, A2, A3, and A4, and here we list the five primary design specifications: FOV, F-number, EFL, TTL, and distortion. The design results including the top-$4$ final lenses designed by OptiNeuro and reference manual design results, and below each lens is labeled with the corresponding Avg RMS spot radius (unit is ${\mu}m$).}  
    \label{fig:fin_res}
\end{figure}

These four design tasks were completed within four days using four NVIDIA GeForce RTX 4090 GPUs. Fig.~\ref{fig:fin_res} presents the design results comprising top-$4$ final lenses generated by OptiNeuro and reference manual designs under four sets of specifications (A1-A4). 
Each lens is labeled below with its Avg RMS spot radius (unit: ${\mu}m$), showing that OptiNeuro's designs achieve Avg RMS radii comparable to manual designs across all specifications (A1: $1.973\sim2.158$ \vs~$2.135$; A2: $2.403\sim3.920$ \vs~$3.068$; A3: $3.530\sim4.061$ \vs~$4.807$; A4: $3.331\sim3.733$ \vs~$4.018$). 
This performance validates OptiNeuro's quasi-human design capabilities through two critical aspects: (1) successful handling of nine-element aspheric systems with $216$ optimization variables, and (2) simultaneous generation of multiple feasible solutions that match human-expert outcomes.

\subsection{Automated design of a glass-plastic hybrid fisheye lens}
\label{sec:2_3}
We validate OptiNeuro's capability to address unprecedented design specifications through an automated design task for a glass-plastic hybrid fisheye lens integrating spherical glass elements (providing high refractive index and thermal stability) with aspheric plastic components (offering lightweight durability and cost efficiency) to achieve ultra-wide-angle imaging. Building upon existing specifications of spherical fisheye lens designs~\cite{pernechele2021telecentric}, we configure three critical parameters: FOV expanded to $200^\circ$ (\vs~conventional $180^\circ$), reduced F-number to $2.1$ (from $3.0$), and shortened TTL from $110mm$ to $21.5mm$. The image sensor adopts a $1/1.8$-inch format with $2.4\text{-}{\mu}m$ pixel pitch. To explore optimal design architecture, we configure six potential Design Forms (DFs): GGPPSPPPP, GGGPSPPPP, GGPGSPPPP, GGPPSGPPP, GG(GG)SPPPP, and GGPPS(GG)PP, where G, P, and S denote glass element, plastic element, and aperture stop respectively. In addition, (GG) denotes two glass elements cemented together. Detailed implementation requirements are provided in \textcolor{blue}{Supplementary Note~6.1}.

\begin{figure}
    \centering
    \includegraphics[width=0.98\linewidth]{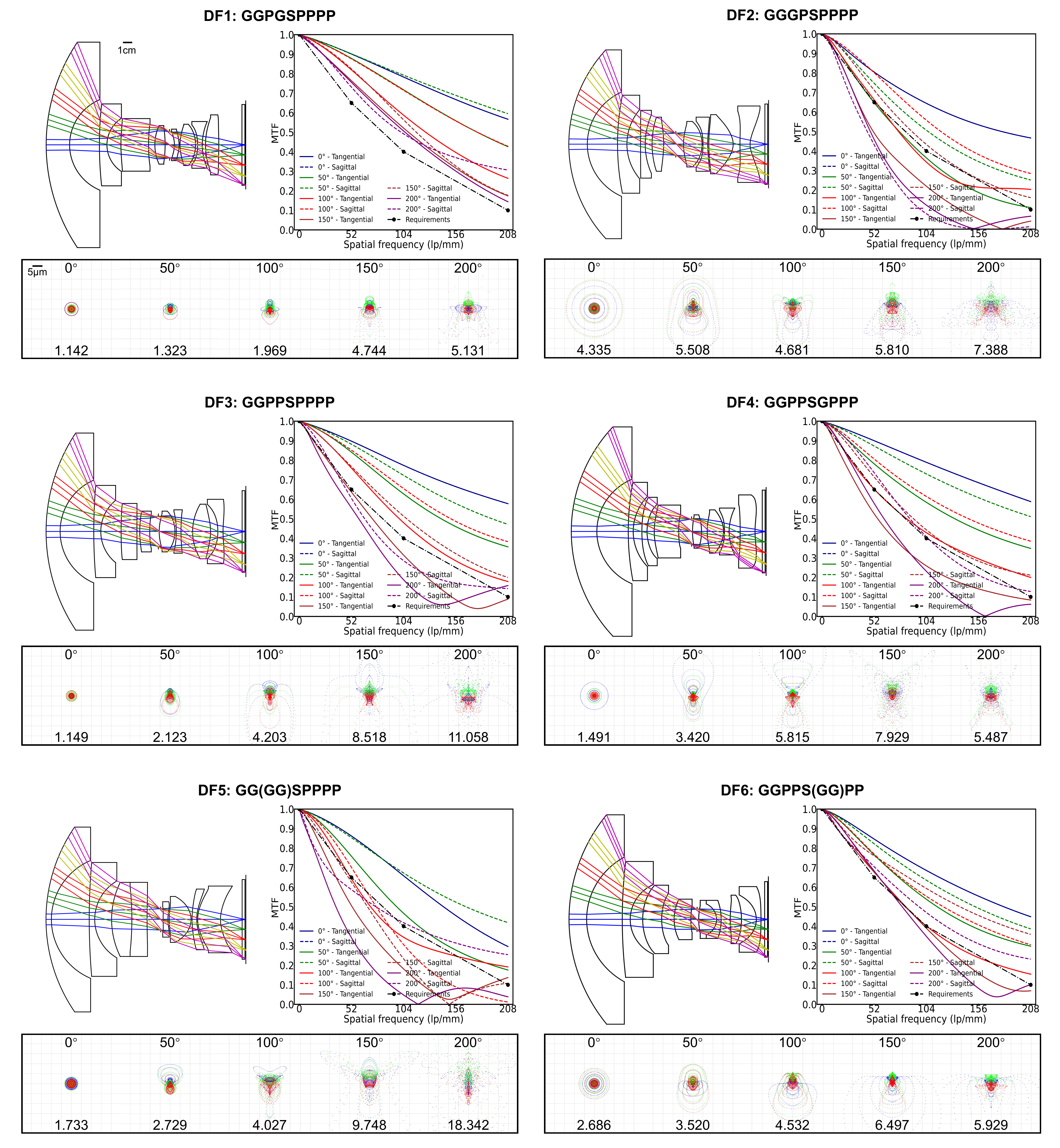}
    \caption{\textbf{Design results of the glass-plastic hybrid fisheye lens.} We configure six potential Design Forms (DFs): GGPPSPPPP, GGGPSPPPP, GGPGSPPPP, GGPPSGPPP, GG(GG)SPPPP, and GGPPS(GG)PP. Each DF is named after its sequence of \textbf{G}lass elements, \textbf{P}lastic elements, and aperture \textbf{S}top. 
    (GG) denotes two glass elements cemented together. We illustrate the optimal solution along with the corresponding spot diagrams and MTF curves for each DF.
    }  
    \label{fig:fisheye}
\end{figure}

This design task was completed under the previous experimental setup within four days using six NVIDIA GeForce RTX 4090 GPUs, where each GPU independently explored feasible solutions within a specific DF. After addressing geometric aberrations, we replaced the image quality metric in the Merit Function from RMS spot radius to RMS wavefront error—an adjustment that optimized the Modulation Transfer Function (MTF) by balancing diffraction-limited performance and energy concentration (\textcolor{blue}{Supplementary Note~6.2}). Fig.~\ref{fig:fisheye} presents the optimal solution for each DF, alongside corresponding spot diagrams and MTF curves. The DF1 (GGPGSPPPP) configuration exhibited a significantly smaller Avg RMS spot radius ($2.862\ {\mu}m$), with MTF values at low- ($52$ lp/mm), mid- ($104$ lp/mm), and high- ($208$ lp/mm) spatial frequencies all meeting design requirements. In summary, OptiNeuro enabled designers to rapidly identify suitable DFs and viable solutions under unprecedented requirements via multi-GPU collaboration.

\section{Discussion}
In this work, we validate the effectiveness of OptiNeuro by designing complex refractive spherical and aspheric lenses—without designing more specialized types such as freeform, reflective, or off-axis lenses. The primary differences in designing these lenses lie in their Merit Functions, while optimization strategies remain consistent overall. Consequently, OptiNeuro can be easily adapted to more specialized lens design tasks in subsequent applications.

It should be noted that while OptiNeuro can autonomously generate high-quality candidate solutions based on input design specifications, it cannot fully replace experienced designers in subsequent analysis steps (\eg, cost estimation, tolerance analysis, and stray light analysis). Additionally, OptiNeuro focuses on optimizing classic image quality metrics—such as spot diagrams and MTF—without accounting for the impact of downstream post-processing algorithms. However, despite these limitations, whether in classical lens design or deep optics paradigms~\cite{zhang2023large, yang2024curriculum, gao2025exploring}, designers can rapidly evaluate, select, and fine-tune based on the high-quality candidates provided by OptiNeuro. This enables them to effectively address manufacturing feasibility requirements or enhance compatibility between lenses and downstream post-processing algorithms.

Although OptiNeuro currently automates the design cycle of complex aspheric lenses within a practical timeframe ($\approx 1$ week), its efficiency still leaves significant room for optimization. While the field lacks standardized public datasets for complex lens systems, OptiNeuro can utilize multi-GPU resources to autonomously construct large-scale lens databases. As the database expands, the system leverages this growing dataset to enable direct fine-tuning of structurally analogous candidates (instead of random initial configurations), substantially reducing development cycles. Furthermore, such datasets facilitate training generative AI models for real-time-compatible lens design~\cite{cote2021deep}. Overall, by automating high-quality lens design workflows, OptiNeuro holds transformative potential for both lens design and the broader optical industry.

\section{Methods}

\subsection{Merit Function in lens design}
\label{method:1}
The main part of designing a lens is to optimize its imaging performance under stringent physical constraints. 
To achieve this goal, a Merit Function (denoted as $MF$) that measures the ``distance'' between the current state and the target state of the lens must be constructed. 
In this work, $MF$ contains two parts: the image quality function ($MF_{img}$) and the physical constraint function ($MF_{phi}$):
\begin{equation}
\label{eq:mfv}
MF=MF_{img}+MF_{phi}=\sqrt{\frac{\sum_{i=1}^{N_{R}}(I^2_i \times W^{I}_i)}{\sum_{i=1}^{N_{R}} W^{I}_i}}+\sqrt{\frac{\sum_{j=1}^{N_{P}}(P^2_j \times W^{P}_j)}{\sum_{j=1}^{N_{P}} W^{P}_j}}. 
\end{equation} 
Here, $N_{R}$ is the number of sampled rays, $I_i$ denotes the aberration (geometric aberration or wave aberration) represented by the $i_{th}$ sampled ray, and $W^{I}_i$ is the corresponding weight; $N_{P}$ is the number of physical quantities that need to be constrained, $P_j$ denotes the deviation of the $j_{th}$ physical quantity's actual value from the target value, and $W^{P}_j$ is the corresponding weight. 
Based on the high-precision ray tracing technology (\textcolor{blue}{Supplementary Note~2.1}) that simulates the propagation of light through lens systems, evaluation of $MF$ is divided into two steps:
(1) Determining the accurate entrance pupil positions of off-axis fields through iterative ray tracing, also known as ray aiming  (\textcolor{blue}{Supplementary Note~2.2}).
(2) Sampling $N_{R}$ rays within the entrance pupils based on the Gaussian Quadrature (GQ) algorithm (\textcolor{blue}{Supplementary Note~2.3}) and then tracing forward to the image surface to evaluate $I_i$ and $P_j$ in Eq.~\ref{eq:mfv}. In addition, $N_{R}$ and $W^{I}_i$ are determined by the GQ algorithm, and $W^{P}_j$ is determined by the order of magnitude of the corresponding physical quantity~\cite{gao2025exploring}.
The constrained physical quantities involved in this work can be found in \textcolor{blue}{Supplementary Note~2.5}.
To further enhance the evaluation efficiency of $MF$, we deploy OptiNeuro on GPU-based architectures that support parallel ray tracing, which demonstrates significant speed advantages over the CPU-based architectures in multi-lens optimization (\textcolor{blue}{Supplementary Note~2.4}).

\subsection{Automated initialization} 
\label{method:2}
To improve the quality of initial lens structures generated by OptiNeuro, we propose a physics-constrained random initialization strategy, which is divided into the following four steps: (1) The material of each element in a lens is randomly assigned from a predefined material library. 
(2) The on-axis thickness of each lens element and the spacing between adjacent lens elements are randomly assigned within the constraints of the Total Track Length (TTL), which is expressed as
\begin{equation}
\label{eq:method16}
\text{TTL}=\sum_{i=1}^{N_{S}} d_i,~d_i >d^{min}_i
\end{equation}
Here, $N_{S}$ is the number of lens surfaces, $d_i$ is the distance between the surface $i$ and surface $i+1$, which represents the element thickness or the air spacing, and $d^{min}_i$ is the minimum distance between surface $i$ and surface $i+1$. 
(3) The spherical curvatures of the lenses are randomly assigned under the constraints of optical power $\Phi$, which is approximately expressed as:
\begin{equation}
\label{eq:method17}
\Phi=\frac{1}{\text{EFL}}\approx \sum_{i=1}^{N_{S}} (n_i - n_{i-1})c_i,~|c_i| < \Phi.
\end{equation}
Here, EFL is the Effective Focal Length, $n_i$ is the material refractive index between surface $i$ and surface $i+1$, and $c_i$ is the spherical curvature of surface $i$. In particular, $n_0$ represents the refractive index of the material from the object surface to the first surface, usually air. 
(4) Under the constraint of EFL, the curvature parameters of the lens are further optimized using the Simulated Annealing (SA) algorithm~\cite{bertsimas1993simulated} to minimize the paraxial aberrations. 

The randomly generated starting points through the above steps satisfy the constraints of TTL and EFL, with the paraxial image quality showing certain improvements (\textcolor{blue}{Supplementary Note~3.2}).
Based on this random initialization strategy, OptiNeuro rapidly generates a large number of random starting points during the initialization phase and then screens a set of diverse initial structures from them using an Euclidean distance-based method~\cite{gao2025exploring}.

\subsection{Adam for lens optimization}
\label{method:3}
The local optimization algorithm plays a pivotal role in the lens optimization. We employ the Adaptive moment estimation (Adam)~\cite{kingma2014Adam} for automated local optimization in lens design. The primary challenge in applying Adam to lens optimization lies in the initialization of learning rates for different optimization variables, because different variables exhibit varying scales (due to different dimensions) and significant differences in nonlinearity.
Appropriate initial learning rates can approximate highly nonlinear lens optimization as a linear optimization process, thereby significantly simplifying the optimization difficulty. Conversely, inappropriate learning rates often lead to optimization stagnation.
Existing works~\cite{yang2024curriculum,gao2025exploring} address this issue by manually determining the initial learning rates for different variables, which relies on human expertise and lacks scalability for generalization to different lens design tasks. Differently, we propose an improved Adam that estimates the initial learning rate for any variable by testing whether the solution space is approximately linear. Specifically, the Merit Function is expressed as $MF=f(\textbf{x})$, where $\textbf{x}=(x^{(1)},x^{(2)},...,x^{(N_L)})^T$ represents the variable vector of a lens. The gradient vector of the lens is expressed as
\begin{equation}
\label{eq:adam_1}
\nabla f(\textbf{x})=(\frac{\partial f}{\partial x^{(1)}},\frac{\partial f}{\partial x^{(2)}},...,\frac{\partial f}{\partial x^{(N_L)}})^T,
\end{equation}
in which $\partial f / \partial x^{(j)}$ is estimated through numerical differentiation
\begin{equation}
\label{eq:adam_2}
\frac{\partial f}{\partial x^{(j)}} \approx \frac{f(x^{(j)}+\Delta x^{(j)})-f(x^{(j)}-\Delta x^{(j)})}{2 \times\Delta x^{(j)}}.
\end{equation}
Here, $j=1,2,...,N_L$ and $\Delta x^{(j)}$ is used as the initial learning rate for variable $x^{(j)}$. Experimental results demonstrate that compared to manually determining the initial learning rate of lens variables, our method effectively enhances the local optimization performance of Adam in complex lens optimization (\textcolor{blue}{Supplementary Note~3.3}).

\subsection{Random perturbation for escaping the local minima}
\label{method:4}
Local optimization may stagnate due to getting trapped in a local minimum; at this stage, OptiNeuro applies random perturbations to the lens parameters before performing local optimization again to escape this local minimum.
However, for complex lenses, even small perturbations to the parameters may lead to structural anomalies, ray tracing failures, and other issues. 
To address this, we propose a physics-constrained random perturbation strategy. The core idea is to prevent a sharp decline in lens quality while ensuring escape from local minima. 
Specifically, we propose two perturbation strategies:

(1) For a spherical/aspheric lens, OptiNeuro randomly selects a lens element and perturbs its material parameters and the curvature of a lens surface while constraining the optical power to remain constant.
Optical power of the selected lens element is determined as:
\begin{equation}
\label{eq:re1}
\Phi_E=(n - 1)(c_1-c_2)+\frac{dc_1c_2(n-1)^2}{n}.
\end{equation}
Here, $n$ is the material refractive index of this lens element, $c_1$ is the curvature of the first surface, $c_2$ is the curvature of the second surface, and $d$ is the element thickness. The perturbation process is divided into the following two steps: Firstly, the lens material is randomly replaced with another material from the predefined material library, which means $n$ in Eq.~\ref{eq:re1} is updated to $n'$. Secondly, one from $c_1$ and $c_2$ is randomly selected 
to apply a perturbation, maintaining constant $\Phi_E$ in Eq.~\ref{eq:re1}. For example, if $c_1$ is selected for perturbation, it is updated to $c'_1$:
\begin{equation}
\label{eq:re4}
c'_1=\frac{n'\Phi_E+n'(n'-1)c_2}{(n'-1)(n'+dc_2(n'-1))}.
\end{equation}
In summary, this process constrains the optical power of the perturbed lens element to remain unchanged, thereby ensuring that the EFL of the entire lens system does not deviate significantly before and after perturbation. 

(2) For a complex aspheric lens, in addition to selecting the perturbation strategy (1), we can also randomly select the aspheric coefficients from an aspheric surface as the perturbation target, while constraining the surface height variation within a reasonable range.
The height of a standard aspheric surface is defined as a function of the radial distance $r$~\cite{sun2021end}:
\begin{equation}
\label{eq:asp}
    h(r) = \frac{cr^2}{1+\sqrt{1-(1+\kappa)c^2r^2}}+\sum^{N_A}_{i=2}a_{2i}r^{2i},
\end{equation}
where $c$ denotes the curvature, $\kappa$ is the conic coefficient, $a_{2i}$'s are higher-order coefficients, and $N_A$ defines the highest-order aspheric coefficient. For a standard aspheric lens surface, $N_A$ is typically set to $8$.
OptiNeuro introduces random perturbations $\Delta a_{2i}$'s to $a_{2i}$'s, where the perturbation terms satisfy the following constraints:
\begin{equation}
\label{eq:xianxing}
    \left\{
    \begin{aligned}
        \sum^{N_A}_{i=2}\Delta a_{2i}r_{1}^{2i}&=\Delta h(r_{1}) \\
        \sum^{N_A}_{i=2}\Delta a_{2i}r_{2}^{2i}&=\Delta h(r_{2}) \\
        &\vdots  \\
        \sum^{N_A}_{i=2}\Delta a_{2i}r_{N_{D}}^{2i}&=\Delta h(r_{N_{D}}) ,
    \end{aligned}
    \right.
\end{equation}
where $N_{D}=N_A-1$, $r_{N_{D}}$ is the maximum radial distance and $r_{j}$ is expressed as 
\begin{equation}
\label{eq:res1}
    r_{j} = \frac{j}{N_{D}}r_{N_{D}},\quad j=1,2,...,N_{D}.
\end{equation}
$\Delta h(r_{j})$ in Eq.~\ref{eq:xianxing} represents the randomly generated perturbation magnitude of surface height at radial distance $r_{j}$, and $|\Delta h(r_{j})|<1 \mu m$. Generally, Eq.~\ref{eq:xianxing} has a set of unique solution $\Delta a_{2i}$'s when $N_{D}=N_A-1$.
In summary, constraining $\Delta h(r_{j})$ ensures that the surface shape does not undergo significant changes before and after applying perturbations to the aspheric coefficients. 
Experimental results show that the proposed perturbation strategy effectively helps to escape from local minima, thereby improving the quality of the design results (\textcolor{blue}{Supplementary Note~3.4}).

\bibliography{bib}

\clearpage
\renewcommand{\thesection}{Supplementary Note \arabic{section}}
\renewcommand{\figurename}{Supplementary Figure}
\renewcommand{\tablename}{Supplementary Table}

\section{Introduction to existing automated lens design methods}
Existing automated lens design methods can be categorized into four main types: 

\begin{itemize}
    \item \textbf{Heuristic global search methods.} Heuristic algorithms, including Genetic Algorithms (GA), Particle Swarm Optimization (PSO), Simulated Annealing (SA), and Ant Colony Optimization (ACO), are commonly integrated with Damped Least Squares (DLS) optimization or Adam to achieve automated lens design. 
    When designing lenses of moderate complexity ($\leq3$ spherical lens elements)~\cite{guo2019new, tang2019ant}, heuristic methods can directly generate high-quality lens designs from input Design Specifications (DS). However, as the number of spherical lens elements increases, the design complexity grows exponentially. Existing heuristic approaches either require Manual Initialization (MI) to produce high-quality solutions~\cite{sun2021automatic, zhang2020automated} or generate suboptimal lens designs that still require subsequent refinement~\cite{gao2025exploring}. The primary challenge arises from the highly non-convex solution space of complex lenses, which prevents existing heuristic algorithms from autonomously escaping local minima without human expertise guidance.

    \item \textbf{Deep Learning (DL)-based methods.} These works~\cite{cote2019extrapolating,cote2021deep} usually rely on massive training lens data and therefore fail when design specifications have insufficient reference data~\cite{yang2024curriculum}. 
    Due to the absence of comprehensive public databases for complex lenses, such methodologies remain confined to classical design architectures (\eg, Cooke triplets, Double-Gauss lenses), often yielding outputs with insufficient image quality or structural anomalies~\cite{gao2025exploring}.
    
    \item \textbf{Point-by-point optimization methods.} While these approaches~\cite{yang2017automated, zhang2021towards} leverage the inherent design flexibility of lenses to achieve automated design, their applications remain predominantly confined to freeform lens systems. 
    
    \item \textbf{Curriculum Learning (CL)-inspired design methods.} These approaches~\cite{yang2022automatic,yang2024curriculum} typically initiate optimization from low-complexity configurations (small aperture and narrow field of view), gradually increasing the field of view and the size of aperture until specifications are met. When combined with Adam, this enables fully automated design from random initial points. Curriculum learning strategies have achieved breakthroughs in automated design for highly aspheric optical lenses. However, performance remains suboptimal: six-element aspheric optical lenses designed via this method exhibit Average (Avg) RMS spot radius exceeding $17{\mu}m$. 
    Even after relaxing distortion constraints, Avg RMS spot radius remains around $12 {\mu}m$, which significantly exceeds the designed pixel size of $2.65 {\mu}m$. This 
    indicates inherent limitations in curriculum learning for designing high-quality complex aspheric lenses.
\end{itemize}
Table S\ref{tab:existwork} summarizes and compares the design capabilities of these methods. In summary, existing automated design methodologies remain constrained to relatively simple lens configurations or specific design types, while their application to more complex multi-element spherical lens systems or highly aspheric configurations still cannot fully eliminate reliance on human expertise for critical design decisions.

\begin{table*}[!t]
    \footnotesize \tiny
    \centering
    \caption{Summary and comparison of existing automated lens design methods (Abbreviations: DS: Design Specifications; MI: Manual Initialization; DL: Deep Learning; CL: Curriculum Learning).}
    \vspace{0em}
    \label{tab:existwork}
    
\renewcommand{\arraystretch}{1.5}
\setlength{\tabcolsep}{1mm}{

\begin{tabular}{c|cccc}
\hline
\hline
Work & Method & Input & Design scope & Output quality \\
\hline
Light Sci. Appl.~(2017) \cite{yang2017automated} & Point-by-point & DS & $3$-mirror freeform & High \\
\hline
Opt. Express~(2019) \cite{guo2019new} & PSO + DLS & DS & $3$-element spherical & High \\
\hline
AO~(2019) \cite{tang2019ant} & ACO + DLS & DS & $2$-element spherical & High \\
\hline
Opt. Express~(2019) \cite{cote2019extrapolating} & DL & DS & $2$-element spherical & High \\
\hline
Proc. SPIE~(2020) \cite{zhang2020automated} & PSO + SA + DLS & DS + \textcolor{red}{MI} & $8$-element spherical & High \\
\hline
Opt. Express~(2021) \cite{cote2021deep} & DL & DS & $6$-element spherical & \textcolor{red}{Insufficient} \\
\hline
Proc. SPIE~(2021) \cite{sun2021automatic} & PSO + GA + DLS & DS + \textcolor{red}{MI} & $8$-element spherical & High \\
\hline
Light Sci. Appl.~(2021) \cite{zhang2021towards} & Point-by-point & DS & $3$-mirror freeform & High \\
\hline
COSI.~(2022) \cite{yang2022automatic} & CL + Adam  & DS & $4$-element aspheric & \textcolor{red}{Insufficient} \\
\hline
Nat. Commun.~(2024) \cite{yang2024curriculum} & CL + Adam  & DS & $6$-element aspheric & \textcolor{red}{Insufficient} \\
\hline
IEEE T COMPUT IMAG~(2025) \cite{gao2025exploring} & GA + SA + Adam  & DS & $6$-element spherical & \textcolor{red}{Insufficient} \\

\hline
\end{tabular}
}

    \vspace{0em}
\end{table*}

\section{More details of the Merit Function in lens design}
\begin{figure}
    \centering
    \includegraphics[width=0.95\linewidth]{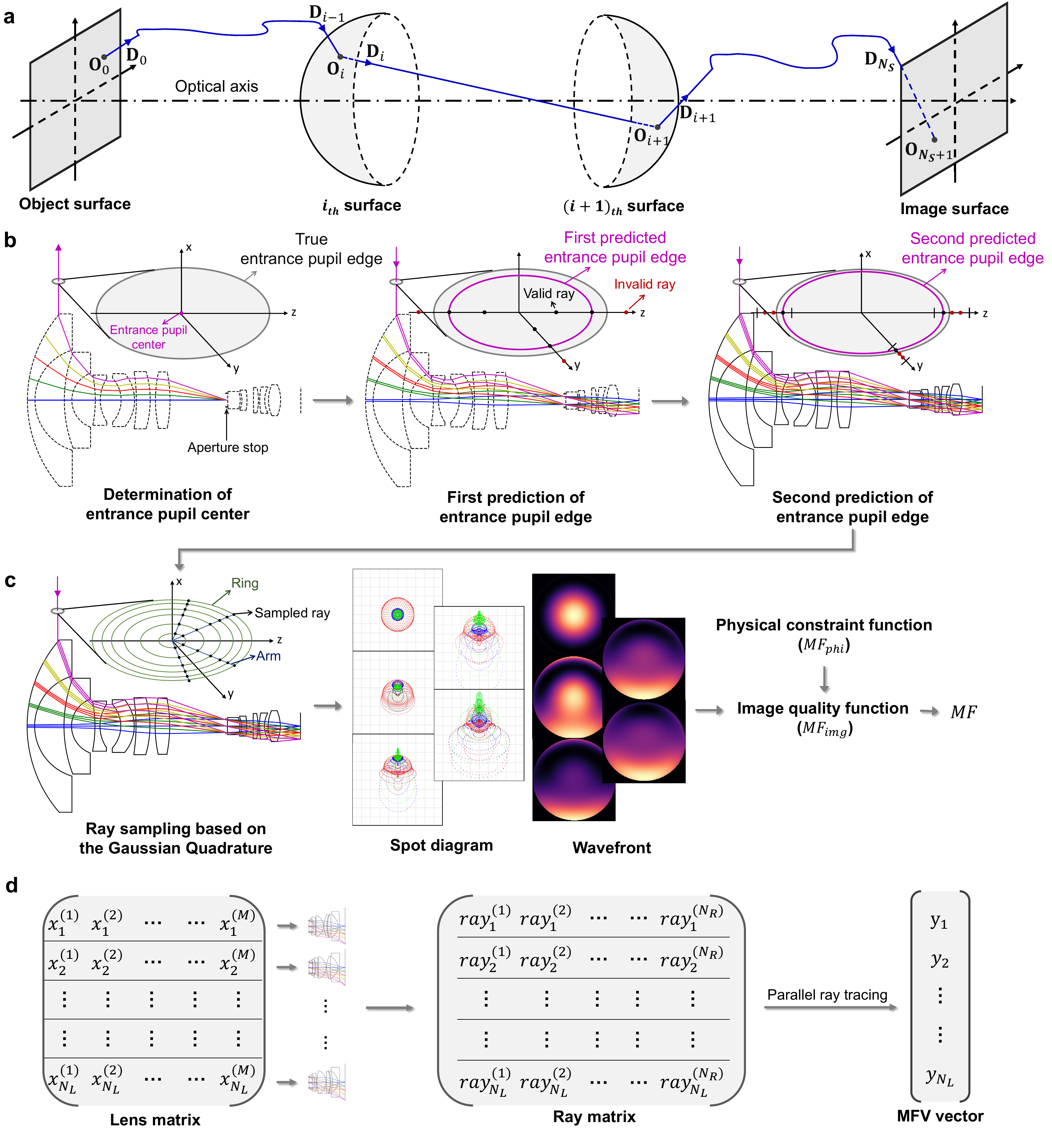}
    \caption{\textbf{Efficient and accurate Merit Function evaluation.} \textbf{a} The general process of ray tracing. This process begins by initializing a ray (blue arrow) at the object surface with position $\mathbf{O}_0$ and direction vector $\mathbf{D}_0$. Given $N_{S}$ lens surfaces (excluding the object and image surfaces), the ray propagates through the system, and the process terminates once the intersection $\mathbf{O}_{N_{S}+1}$ on the image surface is determined. 
    \textbf{b} Three-step ray aiming strategy. Firstly, rays are sampled at the center of the aperture stop and then traced backward to the first lens surface. The approximate entrance pupil center for a specific sampled field angle can be determined. Subsequently, rays are uniformly sampled around the entrance pupil centers along meridional and sagittal directions, and two iterative ray tracings are conducted to determine the entrance pupil edge. 
    \textbf{c} Rays are sampled within the entrance pupils based on the Gaussian Quadrature (GQ) algorithm and then traced forward to the image surface to evaluate the image quality function ($MF_{img}$) and the physical constraint function ($MF_{phi}$), and then further evaluate the Merit Function ($MF$).
    \textbf{d} GPU-accelerated evaluation of Merit Function based on parallel ray tracing. Assuming that there are $N_L$ lenses, where each lens consists of $M$-dimensional lens parameters, and $x^{(j)}_{i}$ represents the $j_{th}$ lens parameter in the $i_{th}$ lens. All lenses form a matrix, which is converted into a ray matrix. Assuming that evaluating the Merit Function of a lens requires tracing $N_R$ rays, a total of $N_L \times N_R$ rays are traced in parallel to obtain the tracing results, and these results are used to compute the MFV (Merit Function Value) vector, in which $y_i$ represents the MFV of the $i_{th}$ lens.}
    \label{fig:mfv}
\end{figure}

\subsection{High-precision ray tracing}
Ray tracing is a fundamental computational technique in lens design. As illustrated in Fig.~\ref{fig:mfv}a, this process begins by initializing a ray at the object surface with position $\mathbf{O}_0$ and direction vector $\mathbf{D}_0$. Given $N_{S}$ lens surfaces (excluding the object and image surfaces), the ray propagates through the system by iteratively performing two steps: (1) Numerically solving for the intersection of the ray and the $i_{th}$ surface based on the concept of a Virtual Conical Surface (VCS) and Newton's method, and updating the ray’s position to $\mathbf{O}_i$. (2) Updating the ray’s direction to $\mathbf{D}_i$ based on Snell’s Law, while incrementing $i$ by one. The process terminates once the intersection $\mathbf{O}_{N_{S}+1}$ on the image surface is determined. Below is a detailed introduction.

The main factor affecting the accuracy of ray tracing lies in solving for the intersection between the ray and the lens surface. The height of a standard aspheric surface~\cite{sun2021end} is defined as a function of the radial distance $r$:
\begin{equation}
\label{eq:res1}
    h(r) = h_0(r)+h_{asp}(r)=\frac{cr^2}{1+\sqrt{1-(1+\kappa)c^2r^2}}+\sum^{N_A}_{i=2}a_{2i}r^{2i},
\end{equation}
where $c$ denotes the curvature, $\kappa$ is the conic coefficient, and $a_{2i}$'s are higher-order coefficients. Given a ray $(\mathbf{O}, \mathbf{D})$, we need to determine $t>0$ such that
\begin{equation}
\label{eq:res2}
   g(\mathbf{O}+t\mathbf{D})=h(r)-z=0.
\end{equation}
We employ a two-stage computation process based on the concept of a Virtual Conical Surface (VCS) and Newton's method to quickly determine $t$. Although Eq.~\ref{eq:res2} generally lacks an analytical solution, an aspheric surface can be decomposed into a combination of a VCS $h_0(r)$ and a higher-order aspheric component $h_{asp}(r)$.
The intersection between the ray and $h_0(r)$ admits an analytical solution, which allows for a coarse initialization $t_0$ by solving $h_0(r) - z = 0$. And $t_0$ can be determined as:
\begin{equation}
\label{eq:method10}
    g_0(\mathbf{O}+t_0\mathbf{D})=h_0(r)-z=At_0^2+Bt_0+C=0,
\end{equation}
where $A=D^2_x+D^2_y+(1+\kappa)D^2_z$, $B=2(O_xD_x+O_yD_y+(1+\kappa)O_zD_z-D_z/c)$ and $C=O^2_x+O^2_y+(1+\kappa)O^2_z-2O_z/c$. We can solve for $t_0$:
\begin{equation}
\label{eq:method11}
t_0 = \frac{ -B \pm \sqrt{B^2 - 4AC} }{2A},B^2 \geq 4AC.
\end{equation}
There are two solutions, in which the solution that makes the initial intersection $(x_0,y_0,z_0)=\mathbf{O}+t_0\mathbf{D}$ closer to the lens surface vertex in the $z$-axis direction is selected. Then $t$ is solved numerically using Newton's method. At iteration $k$+1, $t_{k+1}$ is updated from previous estimate $t_k$ as:
\begin{equation}
\label{eq:method4}
t_{k+1} \leftarrow t_{k} - \frac{g(\mathbf{O}+t_{k}\mathbf{D})}{g'(\mathbf{O}+t_{k}\mathbf{D})},
\end{equation}
where $g'(\mathbf{O}+t_{k}\mathbf{D})=\nabla g \cdot \mathbf{D}$, in which $\nabla g=(2h'(r)x, 2h'(r)y, -1)$, and $h'(r)$ is defined as:
\begin{equation}
\label{eq:method5}
h'(r) = \frac{1+\sqrt{1-(1+\kappa)c^2r^2}-(1+\kappa)c^2r^2/2}{\sqrt{1-(1+\kappa)c^2r^2}(1+\sqrt{1-(1+\kappa)c^2r^2})^2}+\sum^{N_A}_{i=2}a_{2i}ir^{2(i-1)}.
\end{equation}
The termination condition for the iteration is $g(\mathbf{O}+t_{k}\mathbf{D})<1nm$, which is sufficient to ensure the accuracy of ray tracing.
Finally, the refracted direction vector $\mathbf{D}'$ is determined by Snell’s Law:
\begin{equation}
\label{eq:method6}
\mathbf{D}'=\frac{n_1}{n_2}\mathbf{D}+\Bigg(\frac{n_1}{n_2}\cos \big<\textbf{u},\mathbf{D} \big>-\sqrt{1-\Big(\frac{n_1}{n_2}\Big)^2(1-\cos^2\big<\textbf{u},\mathbf{D}\big>)}\Bigg)\textbf{u}.
\end{equation}
Here, $\textbf{u}$ is the normal unit vector of the surface equation, and $n_1$ and $n_2$ are the refractive indices on both sides of the surface, which are determined by the wavelength of the traced ray and the dispersion formula of the material.
$\cos\big<\cdot,\cdot\big>$ is the operation of computing the cosine value of two vectors. In summary, each time the ray passes through a lens surface, the aforementioned process is executed until it reaches the final surface.

Fig.~S\ref{fig:mfv}a compares the proposed method with an existing approach that initializes $t_0$ based on the virtual tangential plane (VTP)~\cite{sun2021end, wang2022differentiable, wu2024mathematical, chen2021optical}. 
Here, $\mathbf{P}_0^{\mathrm{VCS}}$ and $\mathbf{P}_0^{\mathrm{VTP}}$ denote the initial intersection points estimated using the VCS and VTP, respectively, while $\mathbf{P}$ represents the true intersection. The corresponding radial distances are given by $r_0^{\mathrm{VCS}}$ for $\mathbf{P}_0^{\mathrm{VCS}}$, $r_0^{\mathrm{VTP}}$ for $\mathbf{P}_0^{\mathrm{VTP}}$, and $r_{\max}$ for the maximum radial extent of the aspheric surface. There is a non-negligible probability that $r^{VTP}_0>r_{max}$, which results in the failure of Newton's method, whereas 
$r^{VCS}_0\leq r_{max}$ always holds because $r^{VCS}_0$ is limited by $h_0(\rho)$. Additionally, since $h_{asp}(r)$ is typically considered as a small-range perturbation term, $\textbf{P}^{VCS}_0$ is often closer to $\textbf{P}$ than $\textbf{P}^{VTP}_0$, thereby improving the convergence speed of Newton's method. 
In particular, for simple conical surfaces (\ie, $h_{asp}(\rho)=0$), $\textbf{P}^{VCS}_0$ is exactly $\textbf{P}$ and Newton's method is not required. In summary, compared to the VTP, the proposed VCS helps enhance both the accuracy and efficiency of ray tracing.

\subsection{Three-step ray aiming}
It is necessary to determine the accurate entrance pupil positions of off-axis fields due to the pupil aberrations, also known as ray aiming. 
We propose a three-step ray aiming strategy. 
Fig.~\ref{fig:mfv}b illustrates the ray aiming process of a fisheye lens with significant pupil aberration, and provides an enlarged view of the process for determining the entrance pupil position of the edge field.
First, $N_{R}$ rays are sampled at the center of the aperture stop, with incident angles equally spaced between $0^{\circ}$ and $90^{\circ}$, and then traced backward to the first lens surface. According to the principle of optical reversibility, these rays can be regarded as chief rays emitted from the object surface with specific object field angles and passing through the aperture stop center. Therefore, the approximate entrance pupil center for a specific sampled field angle can be determined using linear interpolation methods. Subsequently, $N_{R}$ rays are uniformly sampled around the entrance pupil centers along meridional and sagittal directions, and two iterative ray tracings are conducted to determine the entrance pupil edge. The discrepancy between the second predicted entrance pupil edge and the true entrance pupil edge is small enough to be considered acceptable. 

\subsection{Ray sampling based on the Gaussian Quadrature}
After ray aiming, $N_{R}$ rays are sampled within the entrance pupils based on the Gaussian Quadrature (GQ) algorithm and then traced forward to the image surface to evaluate the Merit Function, as shown in Fig.~\ref{fig:mfv}c.
The GQ algorithm uses a carefully selected and weighted ray set to accurately compute the RMS image quality errors over a uniformly illuminated entrance pupil, which has been proven accurate and requires fewer rays~\cite{forbes1988optical}. The GQ algorithm requires specification of the number of rings $N_{ring}$ and arms $N_{arm}$. Assuming that the lens is rotationally symmetrical and there are $N_{fov}$ sampled fields and $N_{wav}$ sampled wavelengths, $N_{R}$ is calculated as
\begin{equation}
\label{eq:GQ}
N_R=(\frac{N_{fov} \times N_{arm}}{2}-\frac{N_{arm}}{2}+1)\times N_{ring}\times N_{wav}. 
\end{equation} 
In this work, $N_{ring}$ was set to $6$, $N_{arm}$ was set to $8$, $N_{{fov}}$ was set to $5$, and $N_{wav}$ was set to $3$.
Therefore, calculating the Avg RMS spot radius for a single lens using GQ required sampling $306$ rays.

We selected the Average (Avg) RMS spot radius as the image quality metric to validate the effectiveness of the GQ algorithm. Typically, complex aspheric lenses exhibit more irregular structures and entrance pupil positions compared to simple spherical lenses, making it more challenging to predict the RMS spot radii accurately. 
Given the current lack of public datasets for complex aspheric lenses, we generated $5000$ six-element aspheric lenses to construct a test dataset (refer to \textcolor{blue}{Supplementary Note 3.1} for the design specifications).
The reference Avg RMS spot radii of these lenses were computed by Zemax using the hexagonal sampling method, with a distribution range from $2 {\mu}m$ to $20 {\mu}m$.
Hexapolar sampling is a method for distributing light rays in a rotationally symmetric pattern to cover the entrance pupil or exit pupil area. It is characterized by concentric rings of rays around a central reference point. The number of rays increases by $6$ per ring, ensuring uniform angular coverage and minimizing redundancy. 
To ensure computational accuracy, the number of hexapolar rings was set to $64$, and calculating the Avg RMS spot radius using the hexagonal sampling method for a single lens required sampling $187,215$ rays.

Fig.~S\ref{fig:mfv}b compares the predicted Avg RMS spot radius with the reference Avg RMS spot radius, and the unit is ${\mu}m$. 
Due to differences in ray sampling methodologies, the predicted value and the reference value were not entirely identical but exhibited a highly linear relationship-the Pearson correlation coefficient reached $0.99879$. 
This high degree of linear correlation is demonstrated to enable effective lens optimization. Therefore, in this experiment, the GQ algorithm required fewer than $1/600$ of the sampled rays compared to the hexapolar sampling method, yet it remained sufficient to evaluate the magnitude of aberrations (such as the RMS spot radius) in lens optimization.

\subsection{GPU-accelerated evaluation of Merit Function}
In addition to accurate evaluation of the Merit Function, how to evaluate the Merit Functions of multiple lenses in parallel remains a challenge. Existing commercial lens design software typically evaluates the Merit Functions sequentially on CPU-based architectures because a designer usually focuses on optimizing a single lens at a time.
In contrast, OptiNeuro employs multi-lens synchronous optimization to enhance its optimization efficiency.  
To support this and leverage the powerful parallel computing capabilities of modern GPUs, we deployed OptiNeuro on GPU-based architectures. 
As illustrated in Fig.~\ref{fig:mfv}d, there are $N_L$ lenses, where each lens consists of $M$-dimensional lens parameters, and $x^{(j)}_{i}$ represents the $j_{th}$ lens parameter in the $i_{th}$ lens. All lenses form a matrix, which is converted into a ray matrix. Assuming that evaluating the $MF$ of a lens requires tracing $N_R$ rays, a total of $N_L \times N_R$ rays are traced in parallel to obtain the tracing results, and these results are used to evaluate the MFV (Merit Function Value) vector, in which $y_i$ represents the MFV of the $i_{th}$ lens.

We compared the computing time of CPU and GPU architectures across varying numbers of lenses (denoted as $N_L$), using an NVIDIA GeForce RTX 4090 GPU and an Intel Xeon Platinum 8474C CPU, respectively. 
As shown in Fig.~S\ref{fig:mfv}c, when $N_L$ was relatively small (\eg, $N_L=1$), the GPU yielded minimal benefits. However, as $N_L$ scaled to thousands, the GPU demonstrated significant computational advantages, and the computational time gap between the CPU and the GPU gradually widened as $N_L$ grew. At $N_L=10,000$, the GPU took $1.3$ seconds, while the CPU required $15.5$ seconds — over $10×$ slower. 
It should be noted that the computation time on the GPU also slightly increases as $N_L$ increases, which is likely due to limitations in utilization efficiency that prevent full saturation of GPU resources. In summary, our proposed GPU-accelerated method achieves efficient parallel computation of the MFV vector, thereby providing a scalable computational foundation.

\begin{figure}
    \centering
    \includegraphics[width=1.0\linewidth]{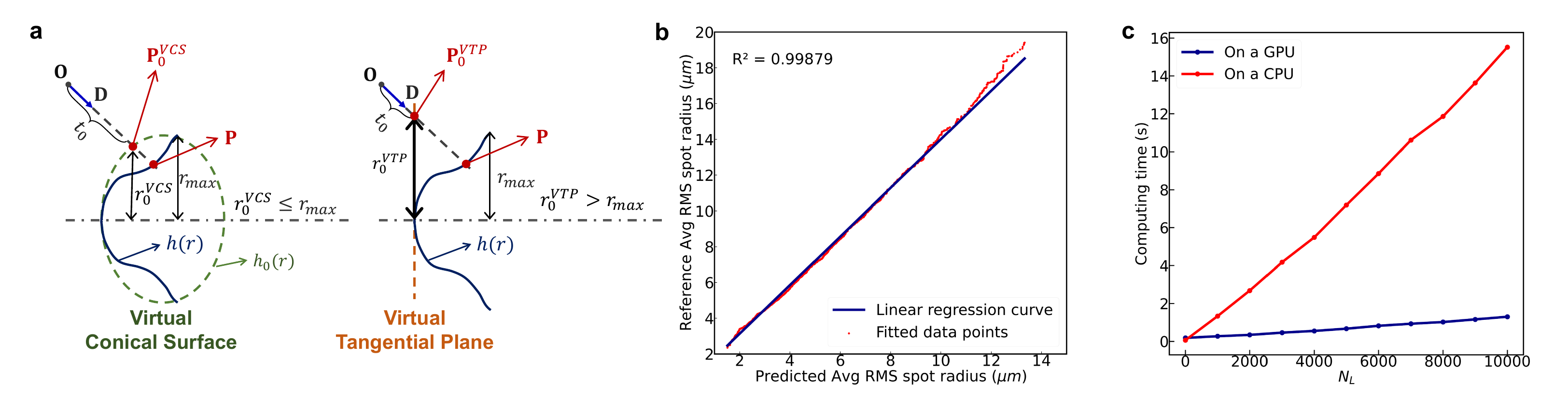}
    \caption{\textbf{Supplementary experimental results about the Merit Function.}  
    \textbf{a} Comparison between the proposed Virtual Conical
    Surface (VCS) with the existing virtual tangential plane (VTP). Here, $\mathbf{P}_0^{\mathrm{VCS}}$ and $\mathbf{P}_0^{\mathrm{VTP}}$ denote the initial intersection points estimated using the VCS and VTP, respectively, while $\mathbf{P}$ represents the true intersection. The corresponding radial distances are given by $r_0^{\mathrm{VCS}}$ for $\mathbf{P}_0^{\mathrm{VCS}}$, $r_0^{\mathrm{VTP}}$ for $\mathbf{P}_0^{\mathrm{VTP}}$, and $r_{\max}$ for the maximum radial extent of the aspheric surface. 
    \textbf{b} Validation of the effectiveness of the GQ algorithm. We generate $5000$ six-element aspheric lenses to construct a test dataset. The reference Avg RMS spot radii of these lenses are computed by Zemax using the hexagonal sampling method, and the predicted Avg RMS spot radii of these lenses are computed using the GQ algorithm. 
    \textbf{c} Verification of GPU acceleration advantages. We compare the computing time of CPU and GPU architectures across varying numbers of lenses (denoted as $N_L$), using an NVIDIA GeForce RTX 4090 GPU and an Intel Xeon Platinum 8474C CPU, respectively.}
    \label{fig:mfv_}
\end{figure}

\subsection{Physical quantities that need to be constrained}
\begin{figure}
    \centering
    \includegraphics[width=1.0\linewidth]{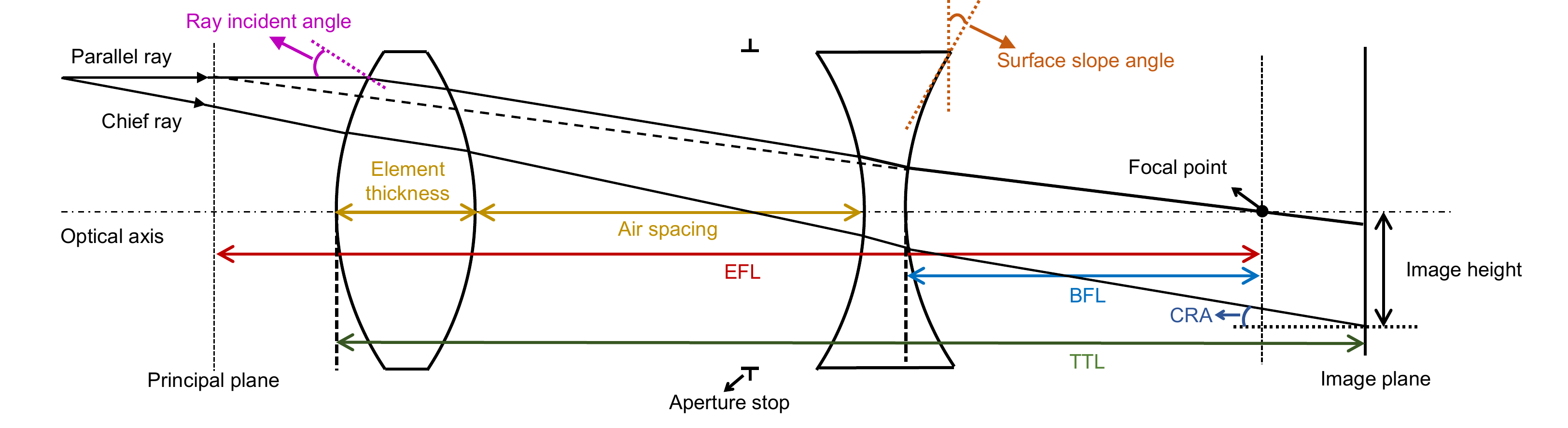}
    \caption{\textbf{Physical quantities that need to be constrained.}}
    \label{fig:phi}
\end{figure}

The physical constraint function measures the deviation of the physical quantities' actual value from the target value or the boundary of the target range. 
Below is a detailed introduction of the physical quantities involved in this work that need to be constrained (as shown in Fig.~S\ref{fig:phi}): 
\begin{itemize}
    \item \textbf{Effective Focal Length (EFL).} A critical parameter in lens design, representing the distance from the principal plane to the focal point when parallel light rays converge. EFL directly determines FOV (shorter EFL = wider FOV) and magnification. Overly short EFL causes excessive distortion and challenges in controlling CRA, degrading edge resolution.

    \item \textbf{Back Focal Length (BFL).} The distance from the last optical surface vertex to the rear focal point when parallel light rays converge. Constraining BFL ensures precise image formation by balancing optical performance (aberration correction, field curvature) with mechanical compatibility (mounting interfaces, system miniaturization). 

    \item \textbf{Total Track Length (TTL).} The distance from the first lens surface to the image plane. Constraining TTL ensures compact system integration while balancing optical performance (aberration control, field curvature) and mechanical feasibility (size, weight), critical for applications like mobile imaging or aerospace, where space and weight are strictly limited.

    \item \textbf{Element thickness.} The physical dimension of an individual lens component along the optical axis. The element thickness must be maintained within appropriate ranges. Excessively thin or thick elements may lead to manufacturing challenges. 

    \item \textbf{Air spacing.} The gaps between optical elements in a lens system. Air spacing between optical elements is critical for balancing aberration correction, mechanical feasibility, and application-specific performance. 

    \item \textbf{Distortion.} Geometric deviations in an optical system that alter the perceived shape, size, or position of objects in an image. Two critical subcategories are F-Tan(Theta) distortion and F-Theta distortion. Constraining distortion ensures geometric accuracy and visual fidelity in optical systems by mitigating curvature-induced shape deformation (\eg, barrel/pincushion distortion), preserving resolution, and enabling precise applications like microscopy or industrial metrology.

    \item \textbf{Ray incident angle.} The angle formed between an incoming light ray and the normal (perpendicular line) at the point where the ray intersects an optical surface. Constraining ray incident angles mitigates aberrations (coma, astigmatism) and optimizes imaging performance by ensuring uniform light distribution and minimizing distortions caused by edge effects or non-uniform ray paths. 

    \item \textbf{Surface slope angle.} The angular deviation of a lens surface from a flat reference surface. Constraining surface slope angles ensures manufacturability, minimizes wavefront errors, and optimizes optical performance. Excessive slopes complicate machining (\eg, diamond turning limitations), induce coma/astigmatism via asymmetric ray paths, and destabilize thermal/mechanical behavior in high-power systems. 

    \item \textbf{Chief Ray Angle (CRA).} The angle between the chief ray (the ray passing through the center of the aperture stop) and the sensor's surface normal at the point where the ray intersects the pixel. This parameter directly impacts light-gathering efficiency, image quality, and compatibility between lenses and image sensors.
\end{itemize}

\section{More details of automated design of the six-element aspheric lens}
In this section, we provide supplementary explanations of the technical details related to the automated design of the six-element aspheric lens in the main paper and supplement experimental validations to demonstrate the effectiveness of the proposed strategy.
\subsection{Design specifications}
The design specifications of the six-element aspheric lens~\cite{yang2024curriculum} are shown in Table~S\ref{tab:d1}. This specification requires a FOV of $80.8^\circ$, an F-number of $2.0$, and a design form of six aspheric plastic elements (denoted as P), with the aperture stop (denoted as S) positioned at the first surface. We revised the original design specifications to ensure more rational and stringent physical constraints. Specifically, in the original design specifications, the Effective Focal Length (EFL) is $5.16mm$, the maximum distortion is $12.8\%$, the image height is $3.83mm$, and the FOV is $80.8^{\circ}$. To constrain lens distortion within acceptable limits, the revised specifications defined the effective focal length as $3.83mm/\tan(80.8^{\circ}/2)=4.5mm$ and set the maximum distortion to $6\%$. In addition, stricter constraints were applied to the element thickness, ray incident angle, and surface slope angle to enhance manufacturing feasibility.

\begin{table}[!t]    
    \caption{Design specifications of the six-element aspheric lens.}
    \vspace{-0em}
    \label{tab:d1}
    
\renewcommand{\arraystretch}{1.2}
\setlength{\tabcolsep}{1mm}{

\begin{tabular}{ccc}
\hline
  & Original specifications  & Revised specifications   \\
\hline
\hline
\multicolumn{1}{c|}{EFL} & $5.16mm$ & $4.50mm$ \\
\multicolumn{1}{c|}{F-Tan(Theta) distortion} & $<12.8\%$ & $<6.0\%$ \\
\multicolumn{1}{c|}{Element thickness} & $>0.25mm$ & $>0.30mm$  \\
\multicolumn{1}{c|}{Ray incident angle}  & $<67^{\circ}$ & $<60^{\circ}$ \\
\multicolumn{1}{c|}{Surface slope angle}  & $<47^{\circ}$ & $<45^{\circ}$ \\

\hline
\multicolumn{1}{c|}{Design form} & \multicolumn{2}{c}{SPPPPPP} \\
\multicolumn{1}{c|}{FOV} & \multicolumn{2}{c}{$80.8^{\circ}$} \\
\multicolumn{1}{c|}{F-number} & \multicolumn{2}{c}{$2.0$} \\
\multicolumn{1}{c|}{Image height} & \multicolumn{2}{c}{$3.83mm$} \\
\multicolumn{1}{c|}{Air spacing} & \multicolumn{2}{c}{$>0.1mm$} \\
\multicolumn{1}{c|}{BFL}  & \multicolumn{2}{c}{$>1mm$} \\
\multicolumn{1}{c|}{TTL}  & \multicolumn{2}{c}{$<6.94mm$} \\
\multicolumn{1}{c|}{Wavelengths}  & \multicolumn{2}{c}{$486nm, 588nm, 656nm$} \\

\hline
\end{tabular}
}

    \vspace{-0em}
\end{table}

\begin{figure}
    \centering
    \includegraphics[width=1.0\linewidth]{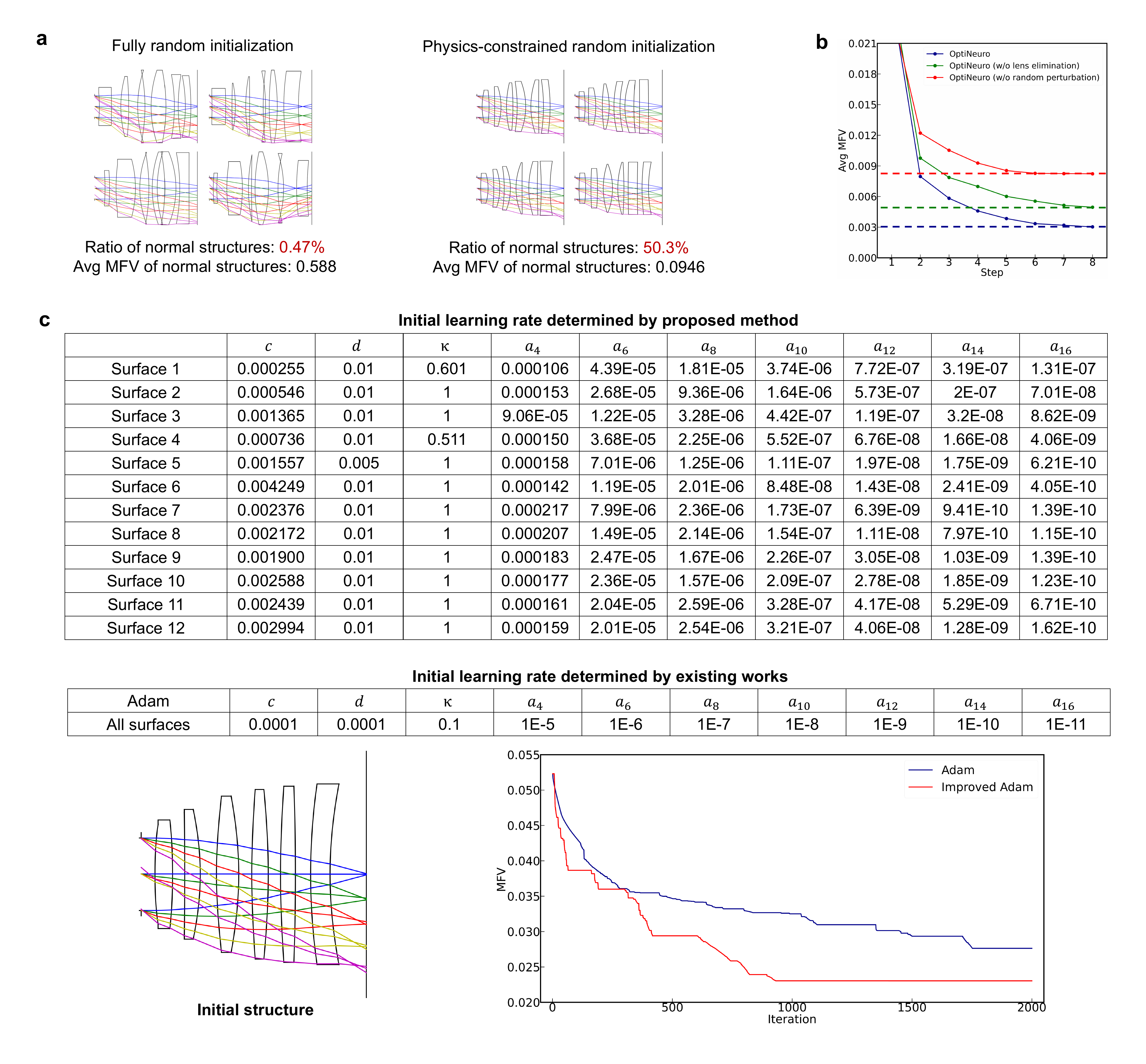}
    \caption{\textbf{Supplementary experimental results to validate the effectiveness of the proposed strategies}  \textbf{a} Comparison of physics-constrained random initialization and fully random initialization.
    \textbf{b} Validation of the effectiveness of the incremental optimization strategy and the random perturbation strategy.
    \textbf{c} Validation of the effectiveness of improved Adam. 
    }
    \label{fig:supp}
\end{figure}

\subsection{Validation of the effectiveness of physics-constrained random initialization strategy}
To validate the effectiveness of the physics-constrained random initialization strategy, we compared it with the fully random initialization method. Specifically, we randomly generated $4000$ initial structures using these two methods, respectively. Among them, the physics-constrained random initialization strategy initialized the curvature and spacing of the lens under the constraints of TTL (Total Track Length) and EFL (Effective Focal Length) while minimizing the paraxial aberrations, while the fully random initialization method had no such constraints. 
Fig.~S\ref{fig:supp}a shows partial normal initial structures generated by the two methods and statistically analyzes the ratio of normal structures as well as the Avg MFV of normal structures. Here, abnormal structures refer to those where rays fail to be traced from the object surface to the image surface. It can be observed that compared to the case without physical constraints, the probability of generating normal structures under physics-constrained conditions was significantly improved, increasing from $0.47\%$ to $50.3\%$. Meanwhile, the Avg MFV of normal structures was also reduced, which indicates an enhancement in the image quality of the initial structures. 

\subsection{Validation of the effectiveness of improved Adam}
We selected an initial structure to perform local optimization and demonstrate the effectiveness of the improved Adam. Unlike existing works that manually determine learning rates for different variables~\cite{yang2024curriculum, gao2025exploring}, we determine the initial learning rate by evaluating the linear ranges of different variables. 
Additionally, when optimization stalls, the learning rates of all current variables will be halved to keep the optimization interval within the linear range. Fig.~S\ref{fig:supp}c shows the initial learning rate determined by our method and the manually specified initial learning rate of existing methods~\cite{yang2024curriculum} under a selected specific initial structure, and the corresponding optimization results. Existing methods set a uniform learning rate for variables of the same type. In contrast, our method automatically assigns different learning rates to each parameter as the linear space ranges of each parameter differ, thus converging faster and enhancing the optimization efficiency. 

\subsection{Validation of the effectiveness of incremental optimization strategy and random perturbation strategy}
To test the effectiveness of the proposed incremental optimization strategy and random perturbation strategy, we conducted two controlled variable experiments: 
(1) The incremental optimization strategy was removed from OptiNeuro, meaning all $121$ parameters (including the basic parameters and higher-order aspheric coefficients) were set as variables at the start, and lens elimination was also not implemented ($N^{(i)}_L$ was fixed to $16$ throughout the entire design process).
It can be observed from Fig.~S\ref{fig:supp}b that the final Avg MFV remained at around $0.005$ after removing the incremental optimization strategy, which was larger than $0.003$ before removal. This phenomenon may arise because the incremental optimization strategy reduced optimization complexity in the early design stage, thereby avoiding premature convergence to local minima regions within high-dimensional non-convex solution spaces. 
(2) The random perturbation strategy was removed from OptiNeuro, meaning no random perturbations were applied to the lens parameters before each local optimization iteration in the refinement process. It can be observed from Fig.~S\ref{fig:supp}b that the final Avg MFV remained at around $0.008$ after removing the random perturbation strategy, which was larger than $0.003$ before removal. This demonstrates that random perturbation effectively helps lenses escape local minimum regions. 
In conclusion, the experimental results demonstrate that both the incremental optimization strategy and the random perturbation strategy significantly enhance the optimization capabilities of OptiNeuro.

\begin{figure}
    \centering
    \includegraphics[width=0.95\linewidth]{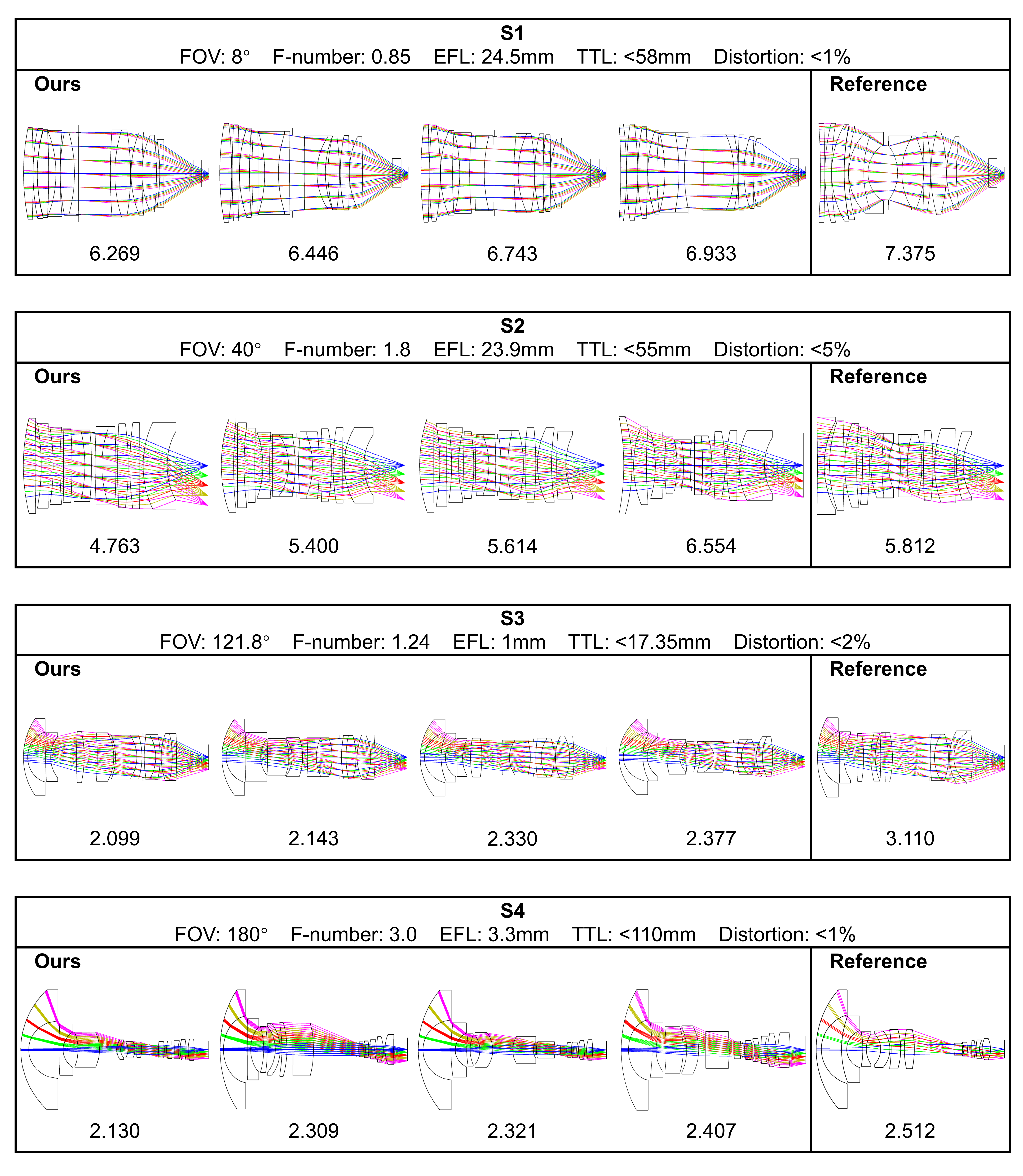}
    \caption{\textbf{First-stage design results of multi-element spherical lenses.} In the first stage, the RMS spot radius was selected as the image quality error. The design results include the top-$4$ final lenses and reference manual design results under four design specifications. Below each lens is labeled with the Avg RMS spot radius.}  
    \label{fig:sph_res_rms}
\end{figure}

\begin{figure}
    \centering
    \includegraphics[width=1.0\linewidth]{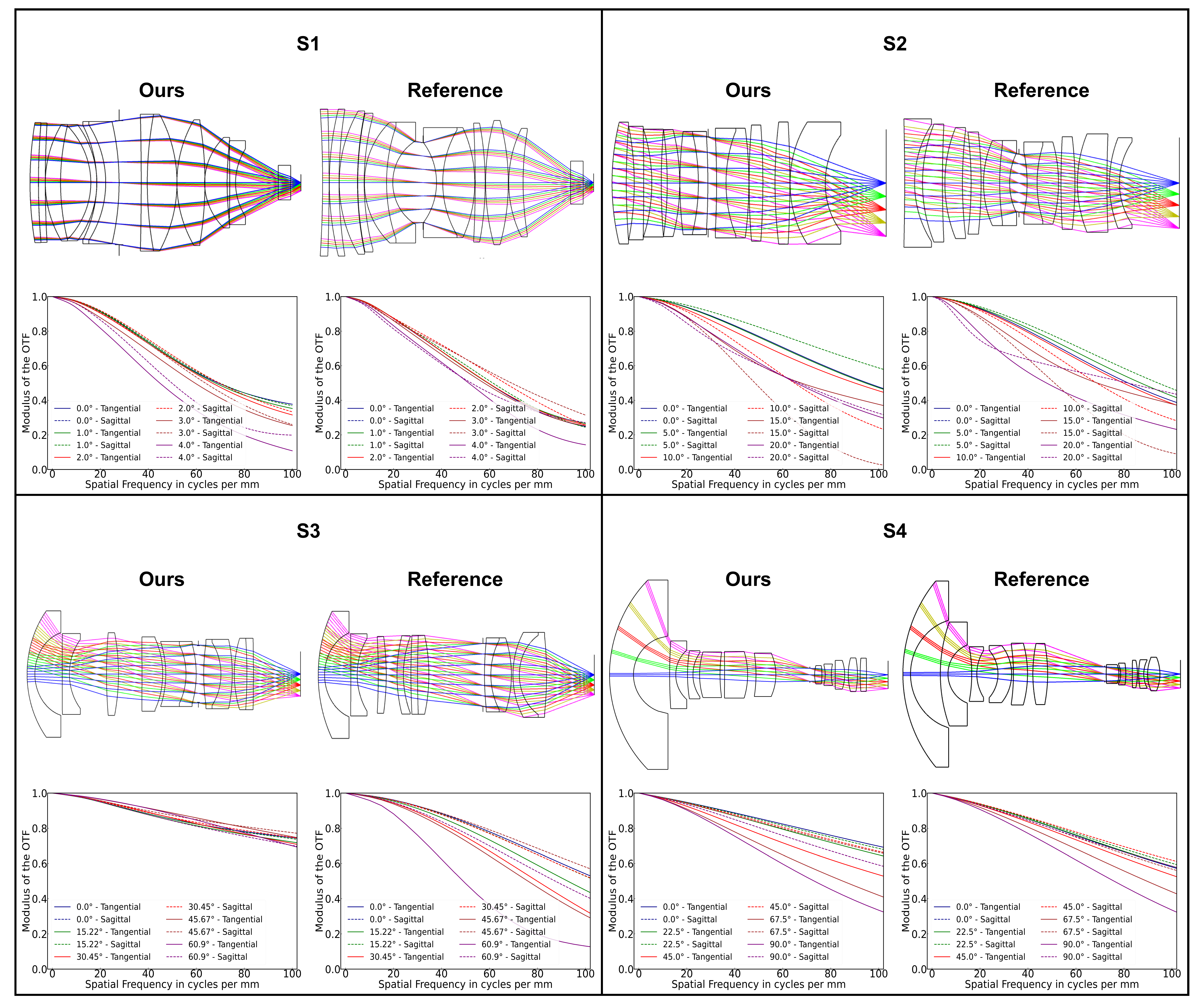}
    \caption{\textbf{Second-stage design results of multi-element spherical lenses.} In the second stage of optimization when geometric aberrations were small, the RMS wavefront error was selected as the image quality metric to further optimize Modulation Transfer Functions (MTF). The design results include the best final lenses and reference manual design results under four design specifications. The corresponding MTF curves are presented below each lens. The maximum frequency is set to $100$ lp/mm.}  
    \label{fig:sph_res_mtf}
\end{figure}

\section{Automated design of multi-element spherical lenses}
\subsection{Design specifications}
Referencing existing public data~\cite{kitahara1999large, nakazawa2000super, pernechele2021telecentric}, we established the design specifications for four multi-element spherical lenses. The design specifications are shown in Table~S\ref{tab:spher}. 
The design specifications of the four spherical lenses are sequentially labeled as S1, S2, S3, and S4. 
These spherical lenses feature varying FOVs (ranging from $8^\circ$ to $180^\circ$) and different F-numbers (from $0.85$ to $3.0$). Comprising more than $10$ glass elements, they are considered complex lenses~\cite{sun2016lens}.
\begin{table}[!t]
    \footnotesize \small
    \caption{Design specifications of four multi-element spherical lenses.}
    \vspace{0em}
    \label{tab:spher}
    
\renewcommand{\arraystretch}{1.2}
\setlength{\tabcolsep}{1mm}{

\begin{tabular}{ccccc}
\hline
  & S1  & S2  & S3  & S4  \\
\hline
\hline
\multicolumn{1}{c|}{FOV} & $8^{\circ}$ & $40^{\circ}$ & $121.8^{\circ}$ & $180^{\circ}$\\
\multicolumn{1}{c|}{F-number} & $0.85$ & $1.8$ & $1.24$ & $3.0$\\
\multicolumn{1}{c|}{EFL} & $24.5mm$ & $23.9mm$ & $1mm$ & $3.3mm$\\
\multicolumn{1}{c|}{TTL}  & $<58mm$ & $<55mm$& $<17.35mm$& $<110mm$\\
\multicolumn{1}{c|}{Distortion} & $<1\%$ & $<5\%$ & $<2\%$ & $<1\%$\\
\multicolumn{1}{c|}{Design form} & \makecell{GGGG(GG)\\S(GG)GGG} &  \makecell{GGGGS\\(GG)GGG} & \makecell{GGGG(GG)\\S(GG)(GG)} &  \makecell{GGGGGG\\SGGGGG} \\
\multicolumn{1}{c|}{BFL}  & $>14mm$ & $>9.5mm$ & $>3mm$ & $>8mm$\\
\multicolumn{1}{c|}{Element thickness} & $>1.3mm$  & $>2mm$ & $>0.4mm$ & $>2mm$\\
\multicolumn{1}{c|}{Air spacing} & $>0.01mm$ & $>0.5mm$ & $>0.1mm$ & $>1mm$\\
\multicolumn{1}{c|}{Ray incident angle}  & $<60^{\circ}$ & $<60^{\circ}$ & $<60^{\circ}$ & $<60^{\circ}$\\
\multicolumn{1}{c|}{Surface slope angle}  & $<50^{\circ}$ & $<50^{\circ}$ & $<60^{\circ}$ & $<80^{\circ}$\\
\multicolumn{1}{c|}{Wavelengths}  &\makecell{486nm\\588nm\\656nm} &  \makecell{486nm\\588nm\\656nm} &  \makecell{486nm\\588nm\\656nm} & \makecell{500nm\\635nm\\770nm}\\ 
\hline
\end{tabular}
}

    \vspace{0em}
\end{table}

\subsection{Design results}
Unlike aspheric lenses, spherical lenses do not incorporate higher-order aspheric coefficients, resulting in a relatively smaller number of variables. Consequently, the incremental optimization strategy and lens elimination strategy were not employed. Other experimental settings were consistent with the glass-plastic hybrid fisheye lens design tasks in the main paper. Specifically, the automated design process was divided into two stages. In the first stage, the RMS spot radius was selected as the image quality metric. Fig.~S\ref{fig:sph_res_rms} represents the design results of the first stage, including the top-$4$ final lenses and reference manual design results under four design specifications. Below each lens is labeled with the Avg RMS spot radius. It can be observed that the spherical lenses designed by OptiNuuro exhibited Avg RMS radii comparable to that of the reference manual designs.
In the second stage of optimization when geometric aberrations were small, the RMS wavefront error was selected as the image quality metric to further optimize Modulation Transfer Functions (MTF). Fig.~S\ref{fig:sph_res_mtf} shows the optimization results of the second stage, including the best final lenses and reference manual design results under four design specifications. The corresponding MTF curves are presented below each lens. The maximum frequency is set to $100$ lp/mm. It can be observed that OptiNeuro-designed lenses exhibit image quality closely approaching that of manually designed systems after optimizing RMS wavefront error, when evaluated against MTF.

\section{More details of designing nine-element aspheric lenses}
\subsection{Design specifications and manual design results}
\begin{figure}
    \centering
    \includegraphics[width=1.0\linewidth]{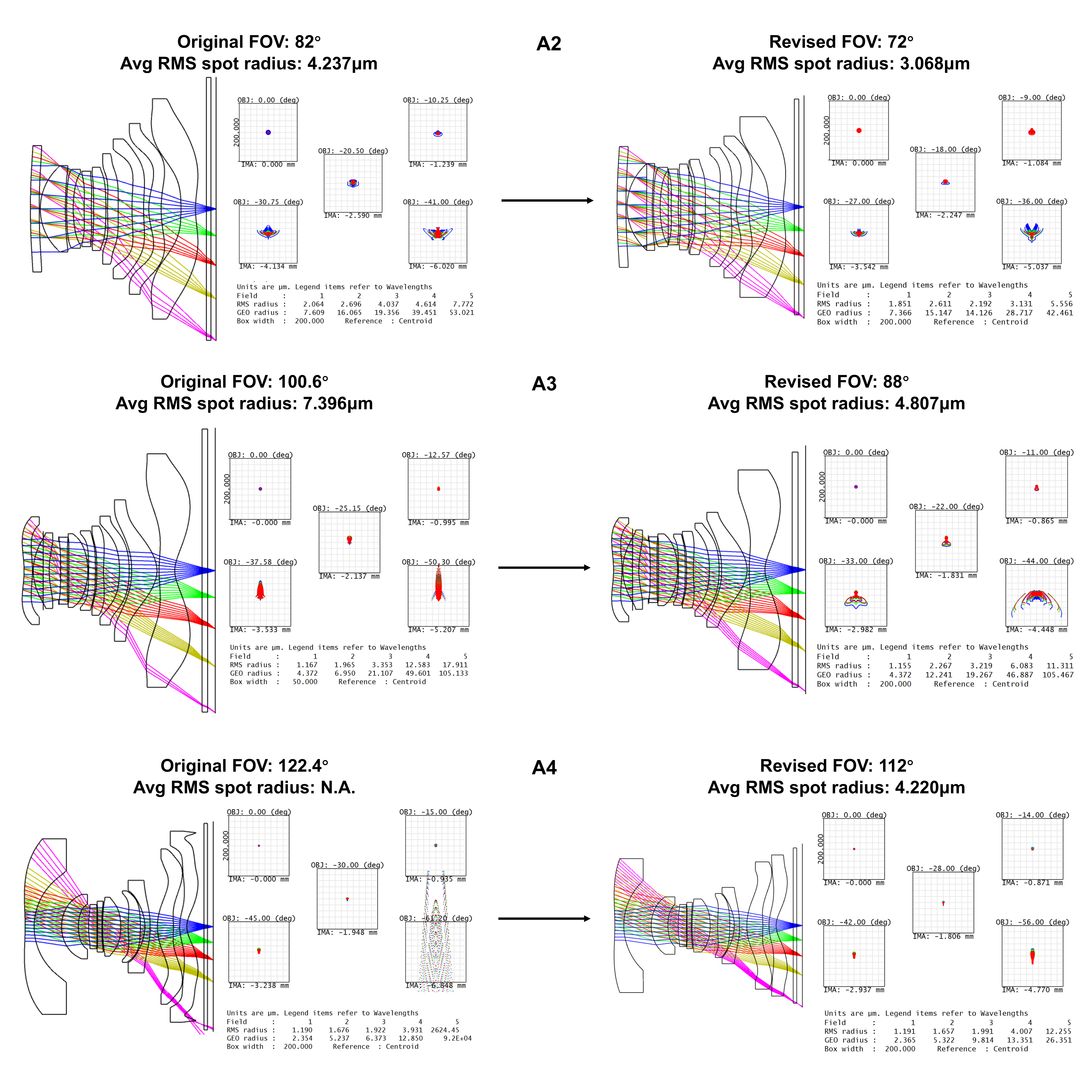}
    \caption{\textbf{Construction of manual design results.} 
    }
    \label{fig:phone9_ref}
\end{figure}

To validate OptiNeuro's capability in automatically designing complex aspheric lenses, we selected four nine-element aspheric lenses in~\cite{chen2022optical, chen2024imaging} as the design target. The design specifications of the four lenses are sequentially labeled as A1, A2, A3, and A4. These patent documents provide artificially designed results under given specifications. However, except for A1, the edge field image quality of the manually designed lenses under A2, A3, and A4 deteriorates sharply, failing to meet practical imaging requirements. Therefore, based on the original specifications, we adjusted the FOVs of these lenses. Fig.~S\ref{fig:phone9_ref} demonstrates that this adjustment significantly improves the average RMS spot radii of these lenses.
The design specifications are shown in Table~S\ref{tab:phone9}. They feature different FOVs and F-numbers but share a short TTL, indicating high compactness of these lenses. Furthermore, the adoption of nine aspheric lenses entails extremely high design freedom, which even poses a notable challenge to human designers.

\begin{table}[!t]
    \caption{Design specifications of four nine-element aspheric lenses.}
    \vspace{0em}
    \label{tab:phone9}
    
\renewcommand{\arraystretch}{1.2}
\setlength{\tabcolsep}{1mm}{

\begin{tabular}{ccccc}
\hline
  & A1  & A2  & A3  & A4  \\
\hline
\hline
\multicolumn{1}{c|}{FOV} & $50^{\circ}$ & $72^{\circ}$ & $88^{\circ}$ & $112^{\circ}$\\
\multicolumn{1}{c|}{F-number} & $2.23$ & $1.7$ & $1.9$ & $2.05$\\
\multicolumn{1}{c|}{EFL} & $6.17mm$ & $6.73mm$ & $4.37mm$ & $3.52mm$\\
\multicolumn{1}{c|}{TTL}  & $<6mm$ & $<8.48mm$& $<7.04mm$& $<11.01mm$\\
\multicolumn{1}{c|}{F-Tan(Theta) distortion} & $<2.5\%$ & $<3\%$ & $<5\%$ & $<8\%$\\
\multicolumn{1}{c|}{Design form} & SPPPPPPPPP & PSPPPPPPPP & PPSPPPPPPP & PPSPPPPPPP\\

\hline
\multicolumn{1}{c|}{CRA}  & \multicolumn{4}{c}{$<35^\circ$}\\
\multicolumn{1}{c|}{BFL}  & \multicolumn{4}{c}{$>0.8mm$}\\
\multicolumn{1}{c|}{Element thickness} & \multicolumn{4}{c}{$>0.3mm$}  \\
\multicolumn{1}{c|}{Air spacing} & \multicolumn{4}{c}{$>0.05mm$} \\
\multicolumn{1}{c|}{Ray incident angle}  & \multicolumn{4}{c}{$<60^{\circ}$} \\
\multicolumn{1}{c|}{Surface slope angle}  & \multicolumn{4}{c}{$<50^{\circ}$} \\
\multicolumn{1}{c|}{Wavelengths}  & \multicolumn{4}{c}{$486nm, 588nm, 656nm$} \\ 
\hline
\end{tabular}
}

    \vspace{0em}
\end{table}

\section{More details of designing a glass-plastic hybrid fisheye lens}

\subsection{Design specifications}
To validate OptiNeuro's capability in efficiently assisting designers to explore feasible solutions under scenarios where optimal design architectures remain unknown, we established an automated design task for a glass-plastic hybrid fisheye lens. The design specifications are shown in Table~S\ref{tab:gpfish}. 
The imaging sensor we applied has a pixel size of $2.4{\mu}m$, which calculates to a Nyquist frequency of $208$lp/mm. Based on this, we set MTF requirements at low ($52$lp/mm), medium ($104$lp/mm), and high ($208$lp/mm) frequencies. 
Based on existing fisheye lens design experience~\cite{fan2019design, pernechele2021telecentric}, we configured $8$ lens elements to balance design costs and imaging requirements. The first two elements are made of glass to effectively control large-angle incident rays, and the aperture stop is placed at the lens center to better balance different aberrations. Based on this premise, we configured six promising design forms and explored the optimal one among them. 

\begin{table}[!t]
    \caption{Design specifications of the glass-plastic hybrid fisheye lens.}
    \vspace{0em}
    \label{tab:gpfish}
    
\renewcommand{\arraystretch}{1.2}
\setlength{\tabcolsep}{1mm}{

\begin{tabular}{cc}
\hline
\hline

\multicolumn{1}{c|}{FOV} & $200^{\circ}$ \\
\multicolumn{1}{c|}{F-number} & $2.1$ \\
\multicolumn{1}{c|}{Image height} & ${4.8mm}$     \\
\multicolumn{1}{c|}{F-Theta distortion} & {$<9\%$}\\
\multicolumn{1}{c|}{Air spacing}     & {$>0.1mm$}  \\
\multicolumn{1}{c|}{Element thickness} & {$>0.5mm$}  \\
\multicolumn{1}{c|}{Wavelengths}      & $435nm, 588nm, 650nm$  \\
\multicolumn{1}{c|}{BFL}  & $>0.8mm$  \\
\multicolumn{1}{c|}{TTL}  & $<21.5mm$ \\
\multicolumn{1}{c|}{Ray incident angle}  & $<70^{\circ}$ \\
\multicolumn{1}{c|}{Surface slope angle}  & $<60^{\circ}$ \\
\multicolumn{1}{c|}{CRA}  & $<38^{\circ}$ \\
\multicolumn{1}{c|}{Relative illumination}  & $>50\%$ \\
\multicolumn{1}{c|}{Sensor format}  & $1/1.1$ inch \\
\multicolumn{1}{c|}{Sensor pixel size}  & $2.4\mu m$ \\

\multicolumn{1}{c|}{IR-Filter material}  & H-K9L \\

\multicolumn{1}{c|}{IR-Filter thickness} & $0.3mm$ \\

\multicolumn{1}{c|}{MTF design requirements} & \makecell{$>65\%$ at $52$ lp/mm\\$>40\%$ at $104$ lp/mm \\$>10\%$ at $208$ lp/mm }\\

\multicolumn{1}{c|}{Design form} & \makecell{DF1: GGPGSPPPP DF2: GGGPSPPPP\\DF3: GGPPSPPPP DF4: GGPPSGPPP \\DF5: GG(GG)SPPPP DF6: GGPPS(GG)PP}\\

\hline
\end{tabular}
}
    \vspace{0em}
\end{table}

\subsection{Balancing spot diagram and MTF}
\begin{figure}
    \centering
    \includegraphics[width=1.0\linewidth]{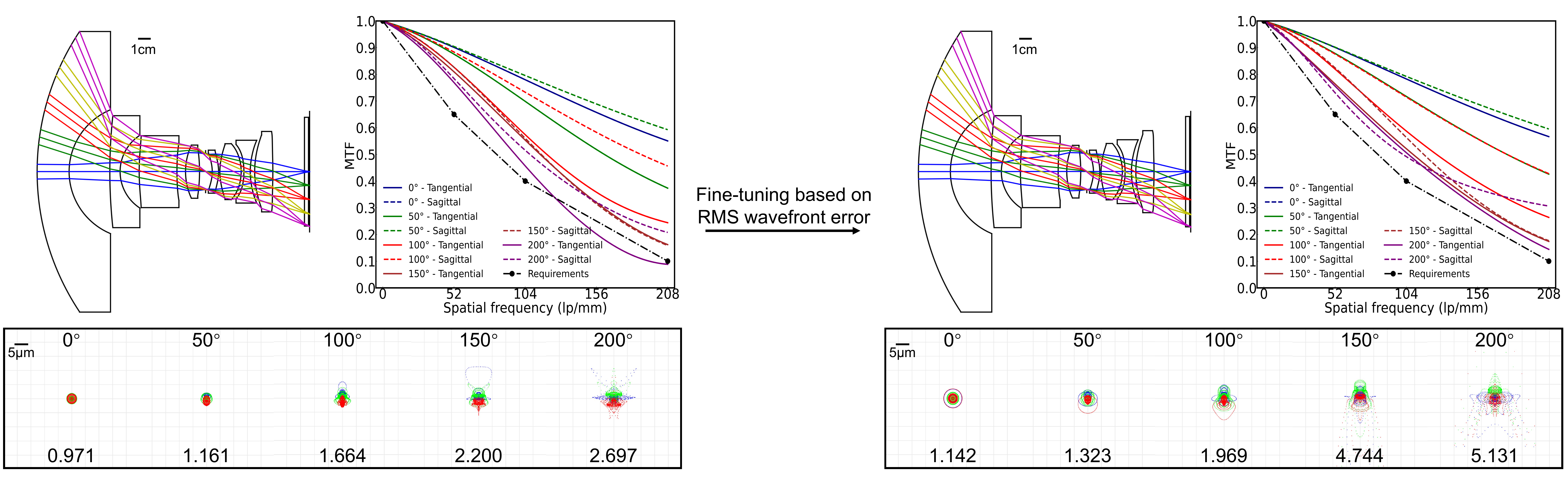}
    \caption{\textbf{RMS wavefront error-based rapid fine-tuning.} 
    }
    \label{fig:wave}
\end{figure}

In lens design, balancing the spot diagram (evaluating geometric aberration-induced energy distribution) and MTF (quantifying frequency-dependent contrast retention) is critical, as they complement geometric and physical optics perspectives: spot diagrams ensure light convergence (energy focus), while MTF preserves detail resolution (contrast maintenance). Imbalance risks either energy dispersion (poor MTF despite small spots) or blurred details (excessive aberrations corrected at MTF’s expense). 
As spot diagrams are applicable to both large and small aberration systems while MTF is more suitable for small aberration evaluation, OptiNeuro first employs RMS spot radius as the image quality metric for automated lens design, followed by RMS wavefront error-based rapid fine-tuning of lens parameters. Fig.~S\ref{fig:wave} demonstrates the spot diagrams and MTF curve variations before/after fine-tuning. By trading off energy concentration (average RMS spot radius increases from $1.739\ {\mu}m$ to $2.862\ {\mu}m$), the lens achieves the required MTF performance, realizing a balance between spot diagram and MTF metrics

\end{document}